\documentclass{PoS} 

\usepackage{heppennames2}

\newcommand{\epem}{\ensuremath{\Pep\Pem}\xspace} 

\newcommand{\ttbar}{\ensuremath{\PQt\PAQt}\xspace} 
\newcommand{\tch}{\ensuremath{\HepProcess{\PQt \to \PQc \PH}}\xspace}
\newcommand{\tcg}{\ensuremath{\HepProcess{\PQt \to \PQc \PGg}}\xspace}
\newcommand{\tcx}{\ensuremath{\HepProcess{\PQt \to \PQc \HepParticle{E}\!\!\!\!\!\!\!\slash\,}}\xspace}
\newcommand{\hbb}{\ensuremath{\HepProcess{\PH \to \PQb \PAQb}}\xspace}

\newcommand{\arxiv}[1]{{\href{https://arxiv.org/abs/#1}{arXiv:{#1}}}}
\newcommand{\http}[1]{{\href{#1}{#1}}}

\graphicspath{{./plots/}}

\usepackage{xspace}

\newcommand{\mylabel}[3]
           {\\
             \vspace*{-#1}
             \hspace*{#2}
{\fontencoding{T1}\fontfamily{phv}\fontseries{b}\selectfont
\scriptsize\hspace*{-1cm}  
  #3 }
\vspace{-\baselineskip}
\vspace*{#1}
           }

\title{On the physics potential of ILC and CLIC}

\ShortTitle{On the physics potential of ILC and CLIC} 

\author{
  \speaker{Aleksander Filip \.Zarnecki} \hspace{\textwidth}
  on behalf of the ILD concept group and the CLICdp collaboration\\
        Faculty of Physics, University of Warsaw\\
        E-mail: \email{filip.zarnecki@fuw.edu.pl}}

\abstract{
The International Linear Collider (ILC) and the Compact Linear
Collider (CLIC) are the two options for a future high-energy,
high-luminosity linear electron-positron collider. Both are expected
to be built in stages, optimised for their physics potential. The main
goals are the precision measurements of Higgs-boson and top-quark
properties as well as direct and indirect searches for new physics
Beyond Standard Model. In my talk I will review some of the latest
results from both ILC and CLIC demonstrating their physics potential,
pointing to similarities and complementarity of both projects.
}

\FullConference{Corfu Summer Institute 2019 "School and Workshops
                on Elementary Particle Physics and Gravity" (CORFU2019)\\
		31 August - 25 September 2019\\
		Corfu, Greece}

\begin{document} 

\section{Introduction}

The European Strategy for Particle Physics is expected to be updated
by May 2020 and will clearly guide the development of the field in the
next decade. 
The research program of the LHC and HL-LHC is unlikely to be
able to solve all the problems of particle physics not addressed in the
Standard Model.    
There is a strong scientific case for an electron-positron collider,
complementary to the LHC, that can study the properties of the Higgs
boson and other particles with unprecedented precision. 
In particular, the worldwide community has been engaged for many
years in two linear collider projects: International Linear Collider
(ILC) which is under consideration to be built in Japan and Compact Linear
Collider (CLIC) at CERN. 
This contribution reviews the physics potential of both ILC and CLIC.
After the two collider projects and the experimental environments have
been described, a personal selection of the physics highlights is
presented, pointing to 
similarities and complementarity of the projects.
The three pillars of the research program for future high energy \epem
colliders will be discussed: Higgs boson studies, top quark physics
and searches for Beyond the Standard Model (BSM) phenomena.
For more information the reader is referred to inputs submitted to the
European Strategy for Particle Physics Update by the ILC \cite{esu_ilc}
and CLIC \cite{esu_clic} collaborations.


\section{Colliders}

\subsection{International Linear Collider}

\begin{figure}[b]
  \begin{center}
    \includegraphics[height=5cm]{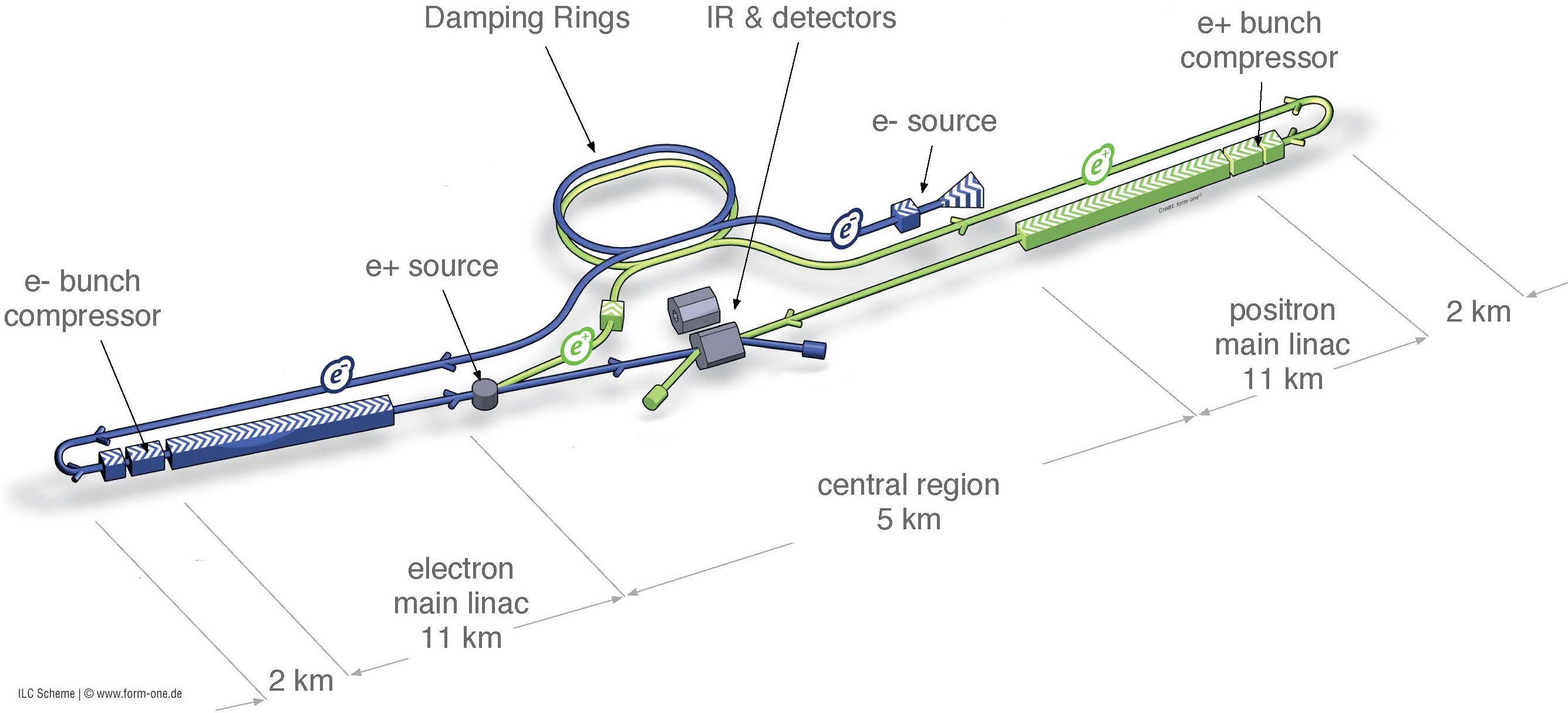}
  \end{center}  
\vspace*{-0.5cm}
  \caption{Schematic layout of the ILC in the 500 GeV configuration,
    as described in \cite{1306.6328}.}
  \label{fig:ilc500}
\end{figure}

\begin{figure}[t]
  \begin{center}
\includegraphics[height=6cm]{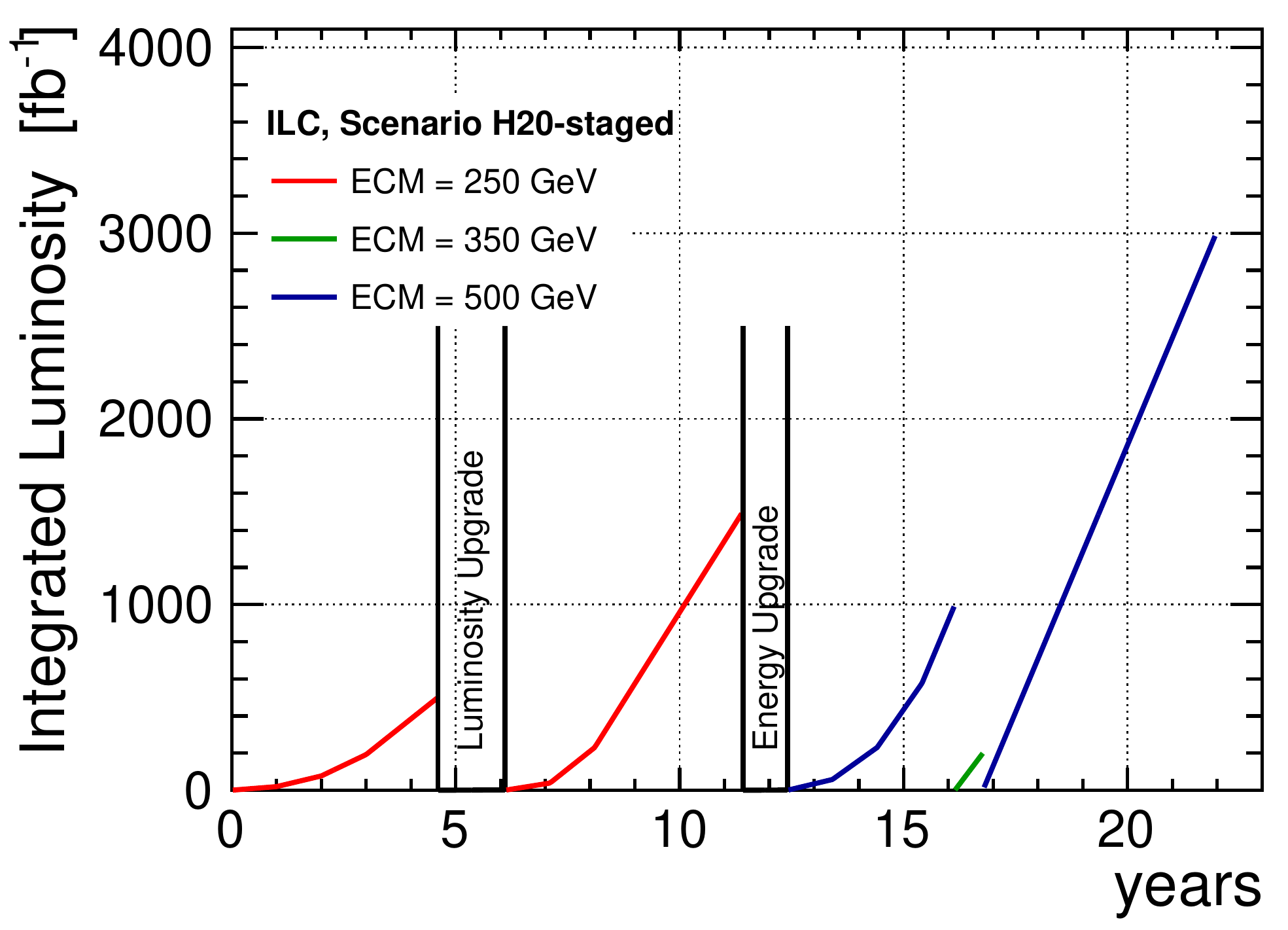}
  \end{center}  
\vspace*{-0.5cm}
  \caption{The nominal 22-year running program for the staged ILC,
           starting operation at 250\,GeV \cite{1903.01629}.}
  \label{fig:h20-staged}
\end{figure}

The International Linear Collider (ILC) project is based on the 
technology of superconducting accelerating cavities.
In the Technical Design Report (TDR) completed in 2013 \cite{1306.6328},
construction of a machine with a centre-of-mass energy of 500 GeV and
a footprint of 31 km was proposed, see Fig.~\ref{fig:ilc500}, with a
possible upgrade to 1 TeV. 
The baseline design includes polarisation for both $e^-$ and $e^+$ beams,
of 80\% and 30\%, respectively.
The discovery of a Higgs Boson with a mass of 125\,GeV motivated
the possibility of reducing initial ILC cost by starting at a
centre-of-mass energy of 250\,GeV \cite{1711.00568} followed by
a 500\,GeV stage and 1\,TeV considered as the possible future upgrade,
should the physics case for such an upgrade be compelling.  
The baseline running scenario for the staged ILC construction is
presented in Fig.~\ref{fig:h20-staged} \cite{1903.01629}.
In the assumed 22-year running period the ILC is expected to deliver
the integrated luminosities of about 2\,ab$^{-1}$ at 250\,GeV and
4\,ab$^{-1}$ at 500\,GeV, with an additional 200\,fb$^{-1}$ collected at
the top-quark pair-production threshold around 350\,GeV.
One has to stress that these integrated luminosities are the same as in
original H-20 proposal for ILC starting at 500\,GeV  \cite{1506.07830}.
ILC is basically ready to be built with all construction issues
verified.
The European XFEL accelerator built with the ILC technology
\cite{1306.6328}, with the nominal energy of 17.5\,GeV. It started
its operation in May 2017, and can be considered the largest ever
accelerator prototype and resulted in full industrialisation
of the cavity production.  
In fact, most of the European XFEL cavities already meet ILC requirements.
Nevertheless, cavity design and production optimisation studies are
still ongoing. 


\subsection{Compact LInear Collider}

\begin{figure}[b]
  \begin{center}
\includegraphics[width=\textwidth]{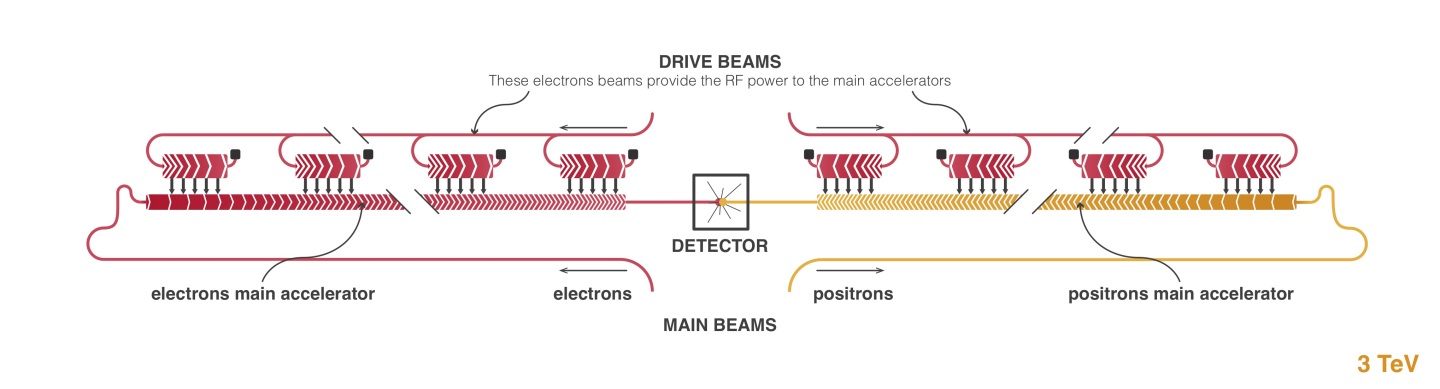}
  \end{center}  
\vspace*{-0.5cm}
  \caption{Schematic layout of the CLIC accelerator \cite{clicdiag}.}
  \label{fig:clic}
\end{figure}

\begin{figure}[t]
  \begin{center}
\includegraphics[width=0.9\textwidth]{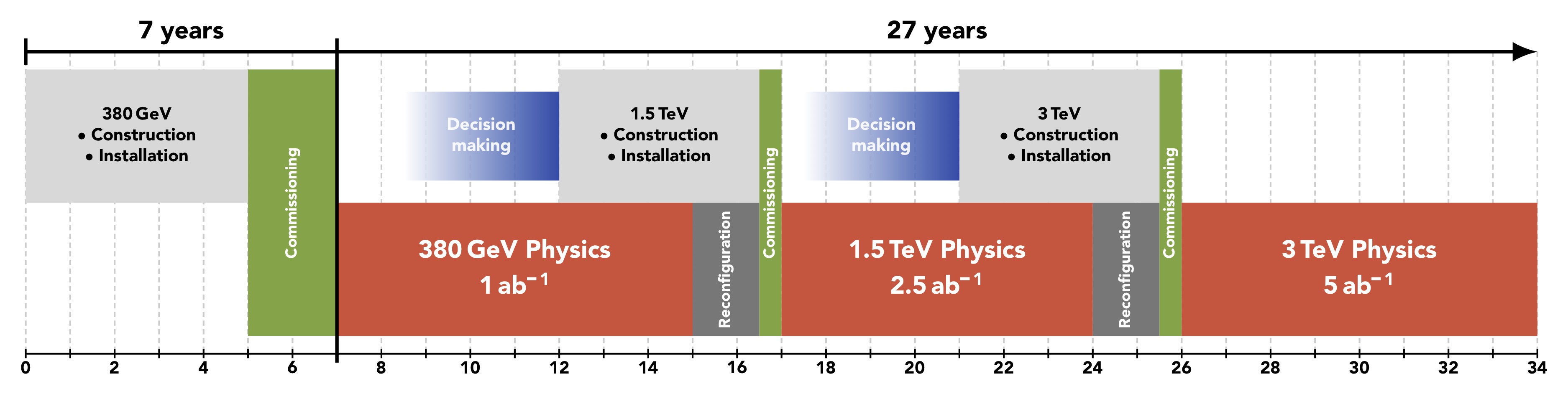}
  \end{center}  
\vspace*{-0.5cm}
  \caption{Proposed construction and running schedule of CLIC \cite{1903.08655}.}
  \label{fig:clic_schedule}
\end{figure}

\begin{figure}[t]
  \begin{center}
\includegraphics[height=5cm]{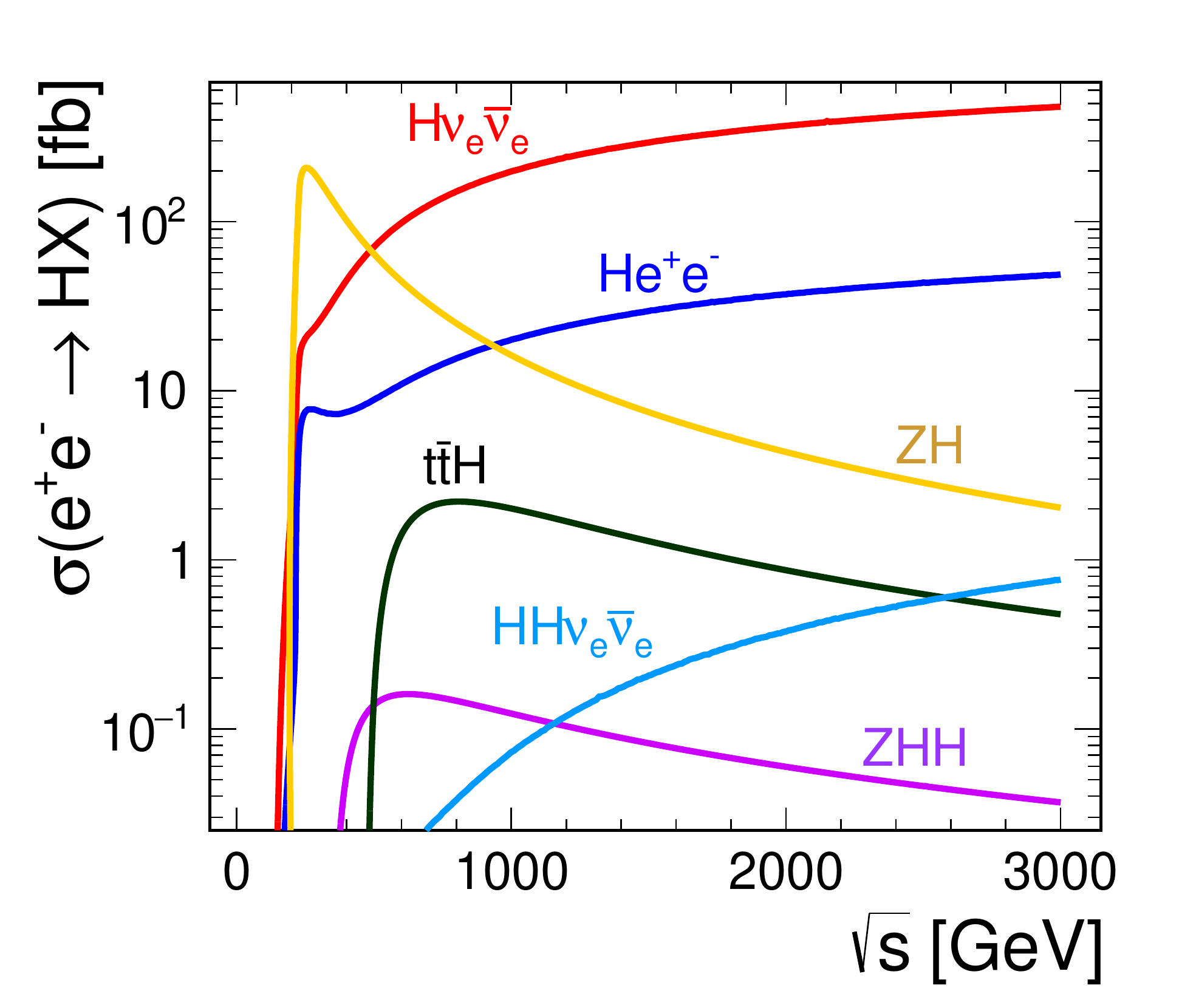}
\includegraphics[height=5cm]{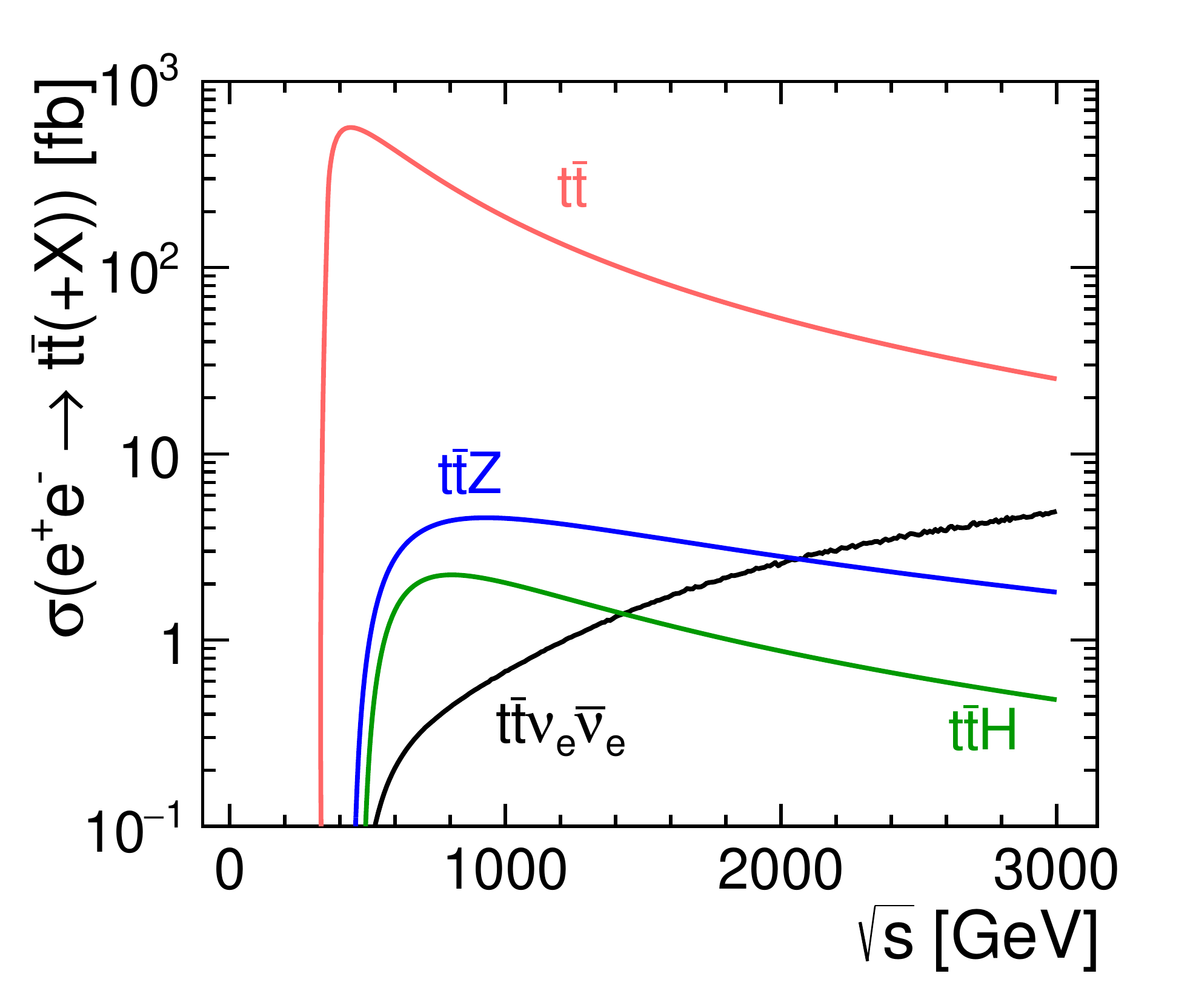}
  \end{center}  
\vspace*{-0.5cm}
  \caption{Production cross section as a function of centre-of-mass
    energy for the main Higgs boson (left) and top quark (right)
    production processes at \epem collider \cite{1812.06018}.}
  \label{fig:htcs}
\end{figure}

The Conceptual Design Report (CDR) for the Compact Linear Collider (CLIC)
was presented in 2012 \cite{Aicheler:2012bya}.
CLIC is based on the two-beam acceleration scheme, required to
generate a high RF gradient of about 100 MV/m, see Fig.~\ref{fig:clic}.
For an optimal exploitation of its physics potential
the implementation plan for CLIC \cite{1903.08655} assumes
construction in three stages, with 7 to 8 years of data
taking at each stage, see Fig.~\ref{fig:clic_schedule}.
The first stage with a footprint of 11 km and an energy
of  380 GeV will focus on precision Standard Model physics and is
optimised for Higgs boson and top-quark measurements.
The plan assumes collecting 1~ab$^{-1}$ at this stage, including
100 fb$^{-1}$ collected at the \ttbar production threshold. 
The second and third construction stages at 1.5\,TeV and 3\,TeV,
with expected integrated luminosities of  2.5\,ab$^{-1}$ and  5\,ab$^{-1}$,
will focus on the searches for BSM phenomena.
However, they will also open possibilities for additional Higgs and
top-quark measurements, such as the double Higgs boson production or
direct determination of the top Yukawa coupling, see Fig~\ref{fig:htcs}.
Only the electron beam polarisation is currently included in the CLIC
baseline design. 
The novel acceleration technology of CLIC required substantial fundamental
research. 
Using the CLIC Test Facility at CERN \cite{Corsini:2017use}, which ended
operation in December 2016, all key elements of the design were verified.

\subsection{Comparison of \epem collider projects}

Table~\ref{tab:lccomp} shows a short comparison of selected
collider parameters of ILC and CLIC.
Both projects have a similar time span.
Thanks to higher expected integrated luminosity, cleaner environment
(thanks to large bunch spacing and size parameters) and positron beam
polarisation, ILC is 
expected to give higher precision at low beam energies, in particular
for Higgs physics (see next section). 
On the other hand, the initial stage energy of CLIC was optimised for
both Higgs boson and top-quark studies and the novel acceleration
technology opens prospects for going beyond the 1\,TeV scale in the
subsequent energy stages.
With different choices of energy stages, running scenarios of ILC and
CLIC are complementary and can give independent constraints on the
Standard Model and different BSM scenarios.
Still, one has to note that the running scenarios will most likely be
revised in the future, depending on new results from HL-LHC and other
experiments. 
\begin{table}[t]
    \begin{center}
      \begin{tabular}{|l|c|c|}
        \hline
                   & ILC & CLIC \\ \hline
    Technology & cold & warm \\
    Acc. gradient & 35 MV/m   & 72/100 MV/m \\
    Initial energy & 250 GeV & 380 GeV \\
    Final energy  & 1 TeV &  3 TeV  \\
    Bunch spacing & 300 ns   &  0.5 ns \\
    Polarisation  & e$^-$ / e$^+$ & e$^-$ \\
    Project timeline & 31 years & 27 years \\
    Total luminosity & 14.2 ab$^{-1}$ & 8.5 ab$^{-1}$ \\ \hline
  \end{tabular}
    \end{center}
    \caption{Comparison of ILC and CLIC projects}
    \label{tab:lccomp}
\end{table}

\begin{figure}[t]
\begin{center}
\includegraphics[height=6cm]{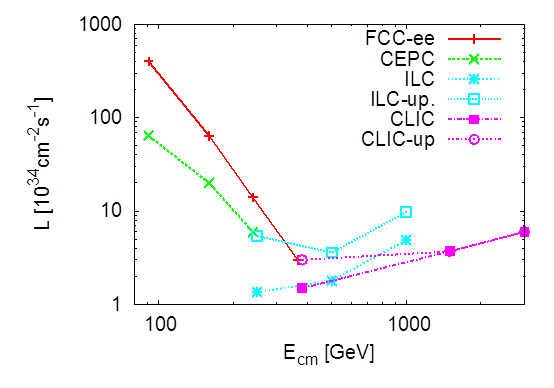}
\end{center}
\vspace*{-0.5cm}
  \caption{The instantaneous luminosities estimated for linear and
    circular \epem collider projects as a function of the
    centre-of-mass energy \cite{Strategy:2019vxc}.
  Two IPs are assumed for the circular colliders FCC-ee and CEPC.}
  \label{fig:lumicomp}
\end{figure}

The instantaneous luminosities
expected for future linear and circular \epem collider projects as a
function of their centre-of-mass energies are presented in
Fig.~\ref{fig:lumicomp}.  
The luminosity expected at circular colliders decreases very rapidly with
increasing energy and it becomes comparable to that of the linear colliders
around the \ttbar threshold.
It is clear that there is no other option for \epem collider at scales
of 500\,GeV or above but the linear collider.
There is also a significant difference in the possible upgrade path of
the two options.
While linear colliders can be upgraded to higher energies by extending
the tunnel, increasing the acceleration gradient or implementing new
acceleration technologies, a significant energy upgrade is not possible
for circular colliders and one can only consider converting them into
hadron machines after the \epem physics program is completed.


\section{Experiments}

  \subsection{Particle Flow concept}

\begin{figure}[t]
\begin{center}
\includegraphics[height=3.8cm]{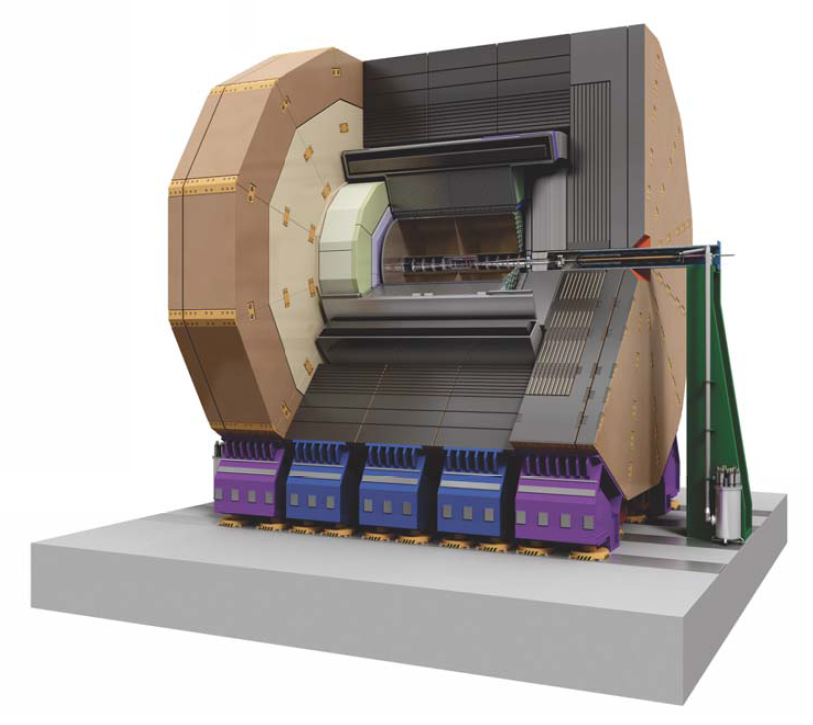}
\includegraphics[height=3.8cm]{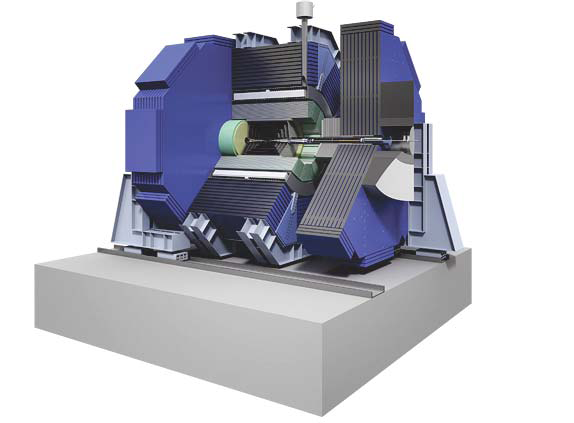}
\includegraphics[height=3.8cm]{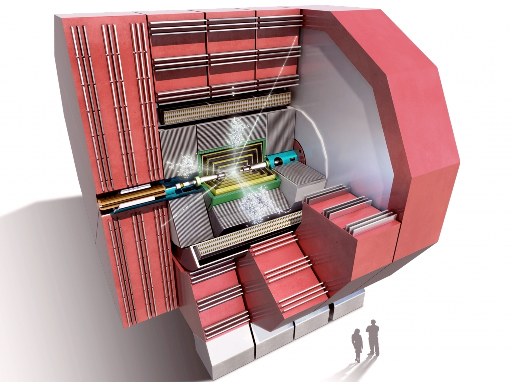}
\end{center}
\vspace*{-0.5cm}
  \caption{Detector concepts for the future linear \epem colliders
    (from left): ILD, SiD and CLICdet.}
  \label{fig:dets}
\end{figure}

Jet energy resolution is crucial for precision physics and background
rejection.
Due to large fluctuations in the hadronic cascade development, jet energy
resolution of calorimeters is limited.
However, significant improvement of the measurement is possible, if we
are able to reconstruct and identify single particles within jets.
The best possible jet energy estimate is then obtained by combining
calorimeter measurements for neutral particles with much more precise
track momentum measurements for the charged ones.
This approach, referred to as Particle Flow Calorimetry
\cite{Thomson:2009rp}, was assumed in the design of the ILC and
CLIC detectors.
Detector designs which were used for the physics potential studies
presented in this contribution are shown in Fig.~\ref{fig:dets}.
Two detector concepts, ILD and SiD, have been developed for ILC
\cite{1306.6329,2003.01116}.
A dedicated detector model, CLICdet, was optimised for full
exploitation of the CLIC physics potential from 380\,GeV to 3\,TeV
\cite{1812.07337}.   
 
\subsection{Detector Requirements and Performance}

\begin{figure}[p]
\begin{center}
 \includegraphics[height=5cm]{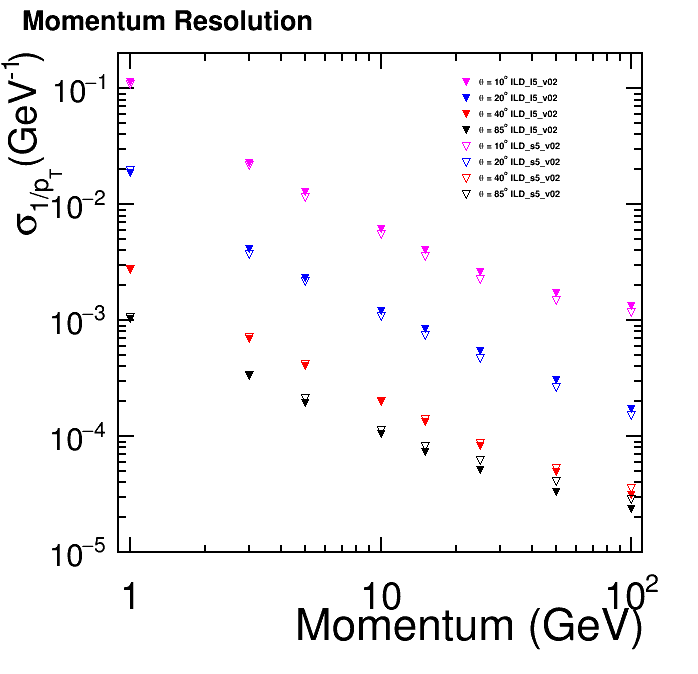}
 \includegraphics[height=5cm]{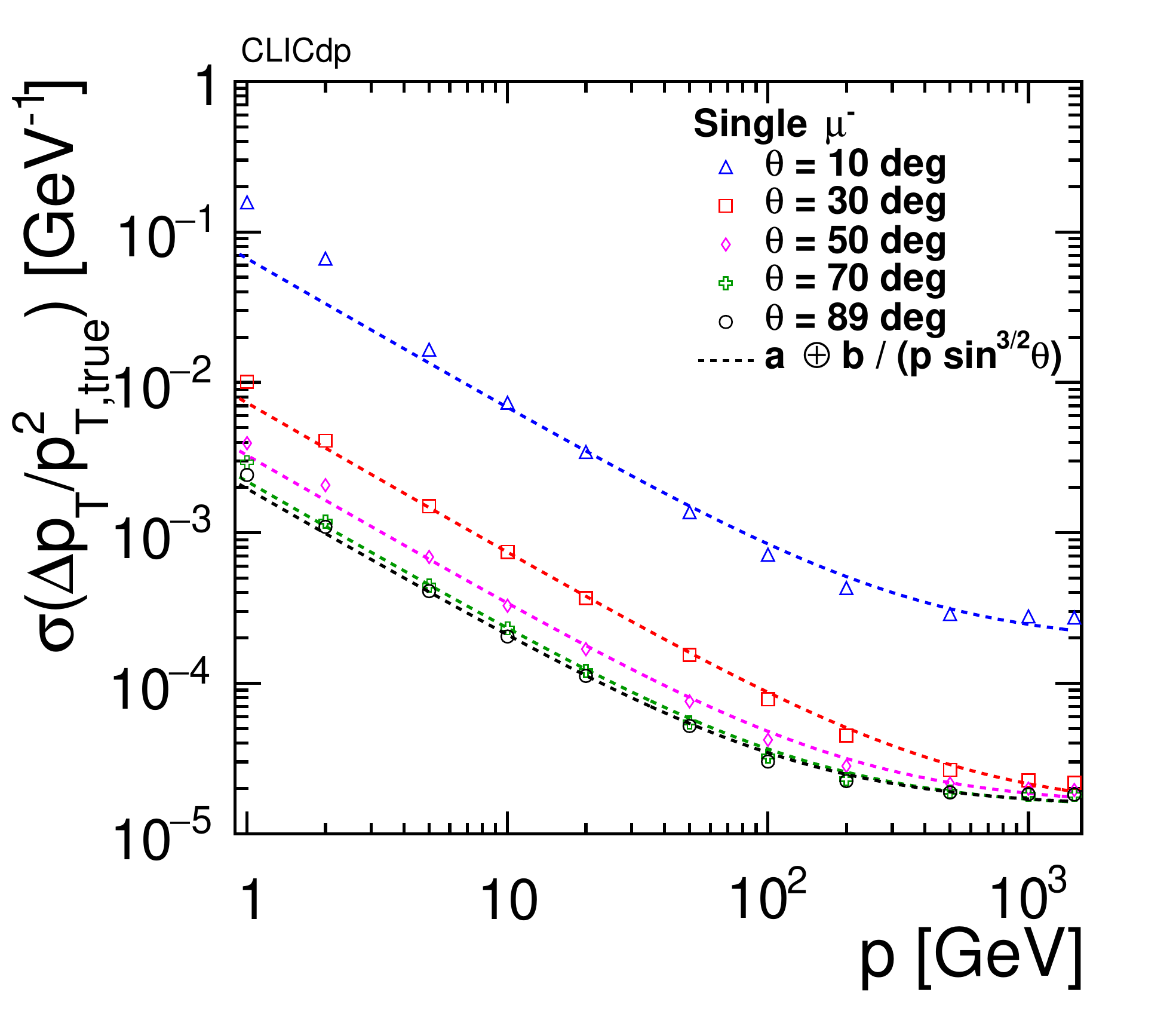}
\end{center}
\vspace*{-0.5cm}
  \caption{Expected resolution in $1/p_{T}$ as a function of the
    track momentum for single muon events, for ILD \cite{2003.01116} (left) and
    CLICdet \cite{1812.07337} (right) detector concepts. }
  \label{fig:ptres}
\end{figure}

\begin{figure}[p]
\begin{center}
\includegraphics[height=5cm]{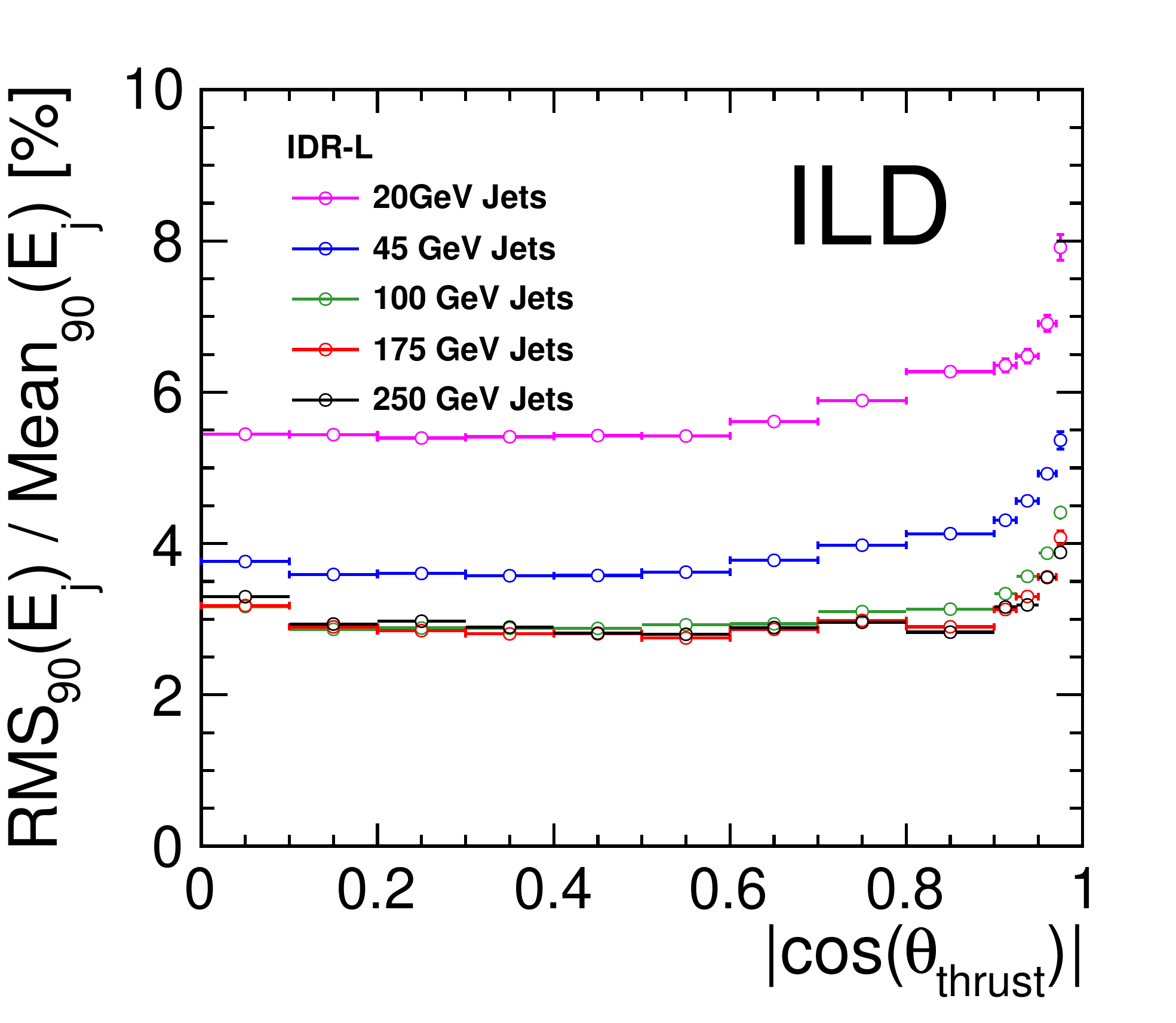}
\includegraphics[height=5cm]{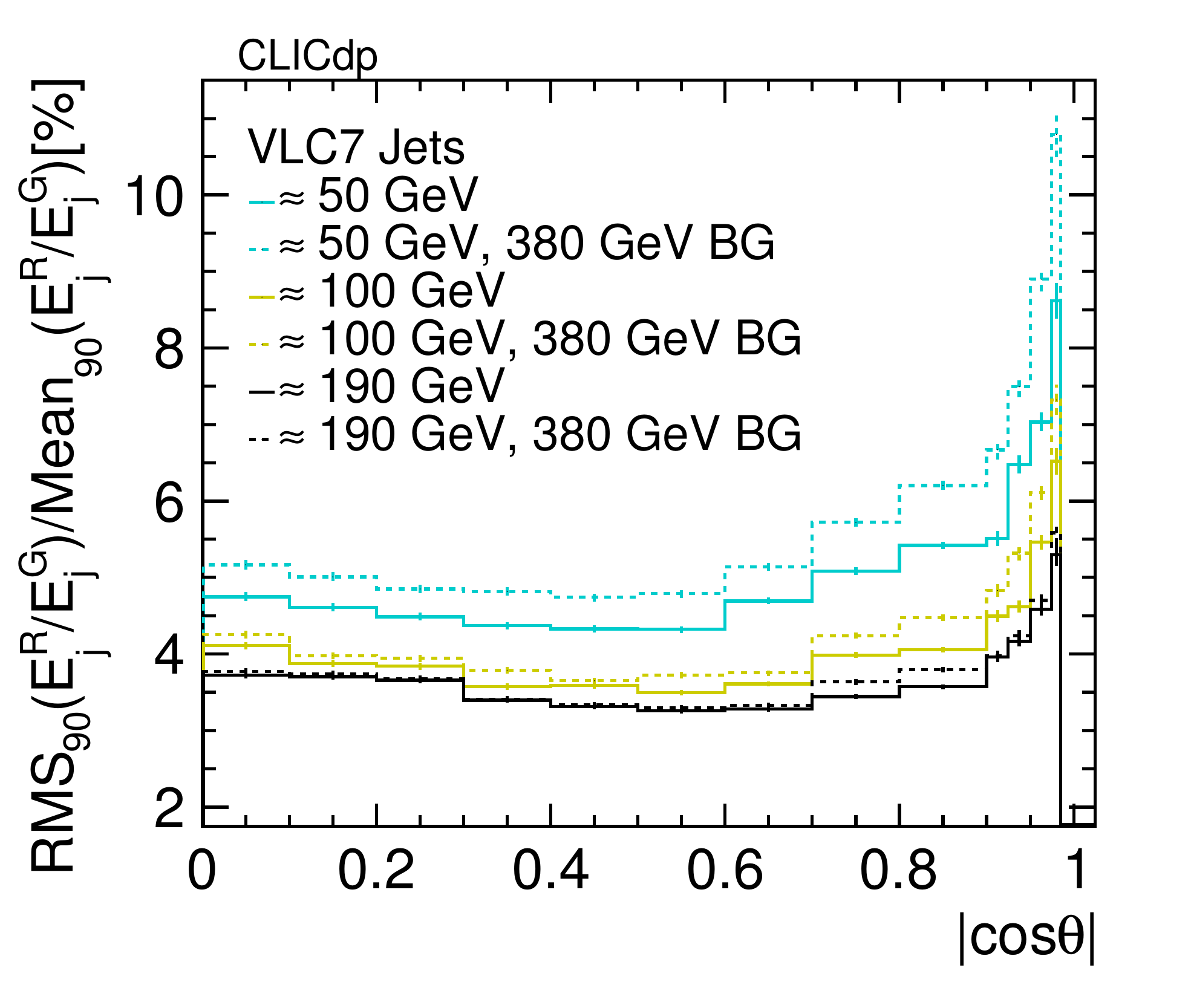}
\end{center}
\vspace*{-0.5cm}
  \caption{Particle flow performance, measured as the energy
    resolution for different jet energies as a function of
    $\cos(\theta)$, for ILD \cite{2003.01116} (left) and CLICdet \cite{1812.07337} (right) detector concepts. 
    The resolution is defined as the rms of the distribution truncated
    so that 90\% of the total jet energy is contained inside the
    distribution.
  }
  \label{fig:jetres}
\end{figure}

Single particle reconstruction and identification is crucial for the 
Particle Flow approach.
It is provided by very high calorimeter granularity, which allows 
efficient matching of calorimeter clusters with the charged particle
tracks and separation of particle clusters inside the detector. 
For the best energy estimate of charged particles, precise momentum
measurement in the wide angular and energy range is required.
Figure~\ref{fig:ptres} shows the expected resolution in $1/p_{T}$
as a function of the track momentum and angle, for single muon events.
For high momentum tracks produced at large angles, simulation studies
indicate that a resolution of 
$\sigma_{1/p_{T}} \sim 2 \cdot 10^{-5}$\,GeV$^{-1}$ can be obtained. 
Expected jet energy resolution based on particle flow reconstruction
is presented in Fig.~\ref{fig:jetres}, as a function of the jet
emission angle for different jet energies.
For high energy jets, resolution of
 $\sigma_{E}/{E} = 3 - 4 $\% 
can be obtained.

\begin{figure}[p]
  \begin{minipage}[b]{0.48\textwidth}
  \hspace{0.2\textwidth}\includegraphics[height=5cm]{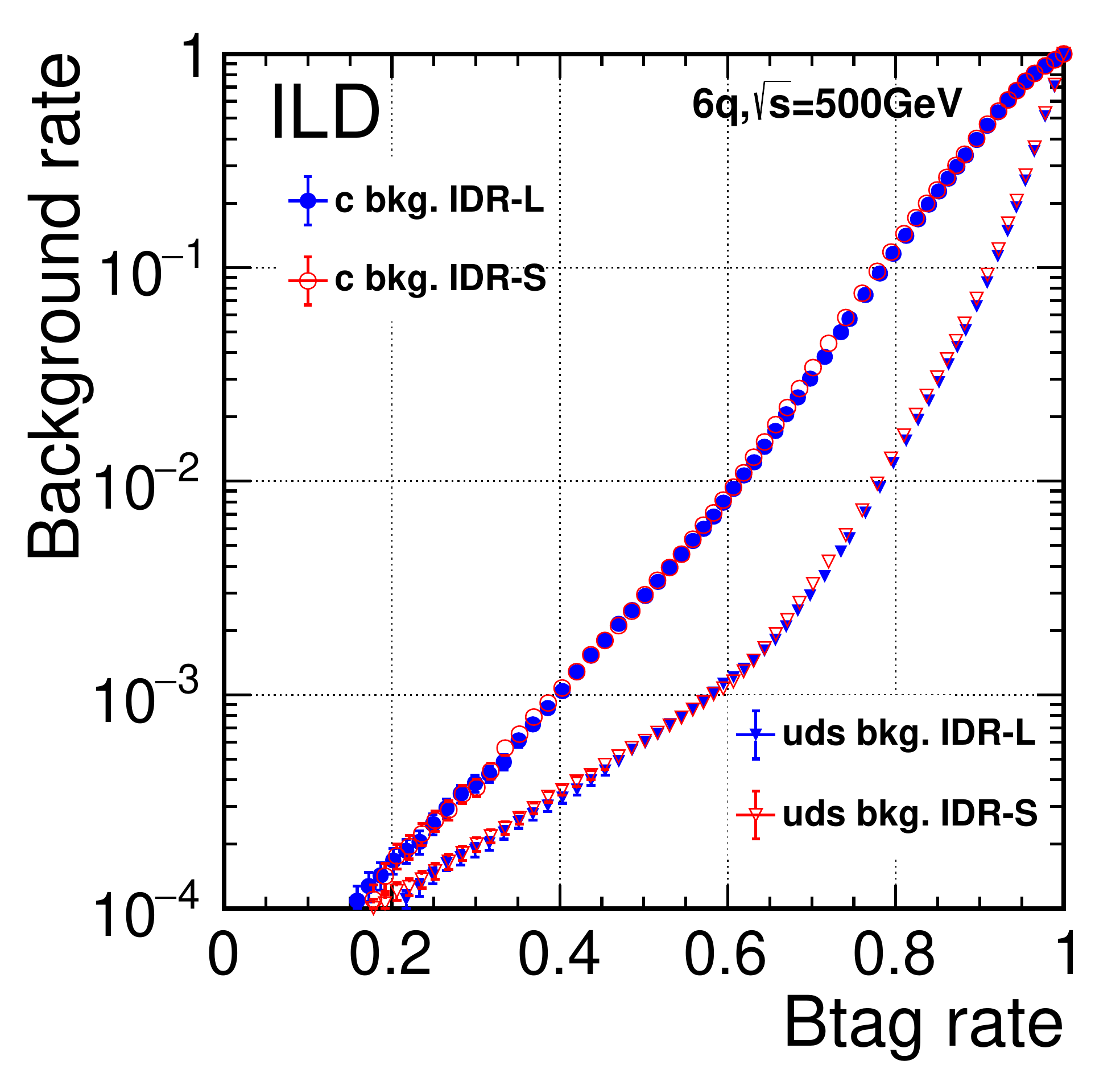}
  \end{minipage}%
  \begin{minipage}[b]{0.48\textwidth}
\includegraphics[width=5.5cm, trim = 0 10.39cm 0 0, clip ]{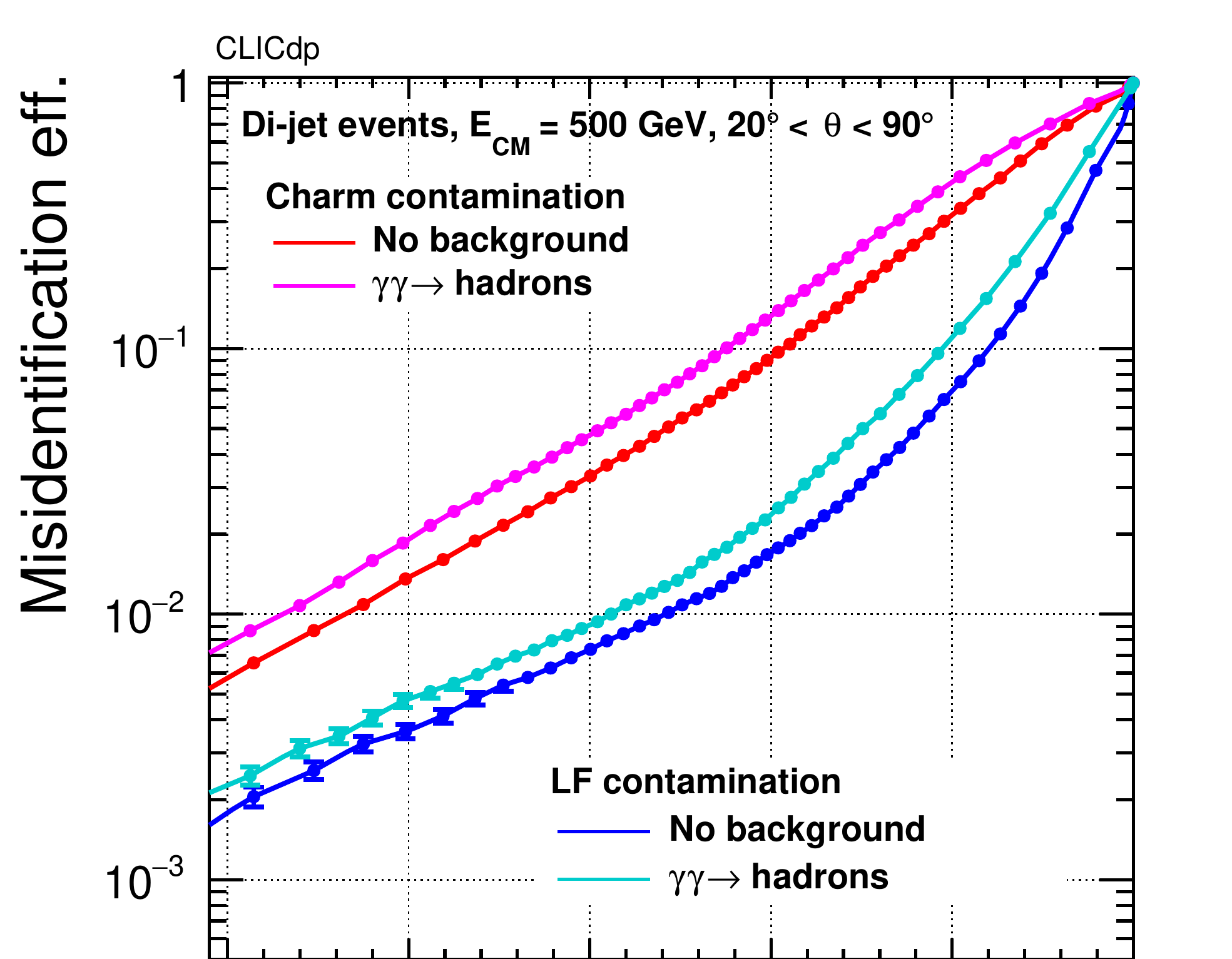}

\hspace{0.85cm}\includegraphics[width=4.6cm, trim = 3.1cm 0.5cm 0 22.9cm , clip ]{plots/Btag_misidEff_background.pdf}
  \end{minipage}
  \caption{Misidentification rates as a function of the b-tagging
    efficiency for c-quark and light flavour (u,d, s) quark jets, expected for
    ILD \cite{2003.01116} (left) and CLICdet \cite{1812.07337} (right)
    detector concepts. } 
  \label{fig:btag}
\end{figure}

With a high precision pixel vertex detector placed very close to the
beam line, precise interaction point determination and very efficient
flavour tagging is possible. 
With the impact parameter resolution for high momentum tracks of
  $\sigma_{d} < 5 \mu m $, bottom and charm quarks can be tagged with
high purity.
Figure~\ref{fig:btag} compares the expected b-tagging
misidentification rates for the ILD and CLICdet detectors.
For quark pair production (including e.g.\ Higgs boson decays)
light-quark background suppression by a factor $\sim10^{-4}$ can be
obtained for a $\PQb\PAQb$ event selection efficiency of 50\%.
Finally, for very good detector hermeticity and
efficient suppression of backgrounds to
processes with missing energy, instrumentation extending
down to a minimum angle of $\theta_{min} \sim 5$~mrad is planned.


\clearpage

\section{Higgs physics}

\subsection{Production processes}

\begin{figure}[t]
\begin{center}
  \includegraphics[height=3cm]{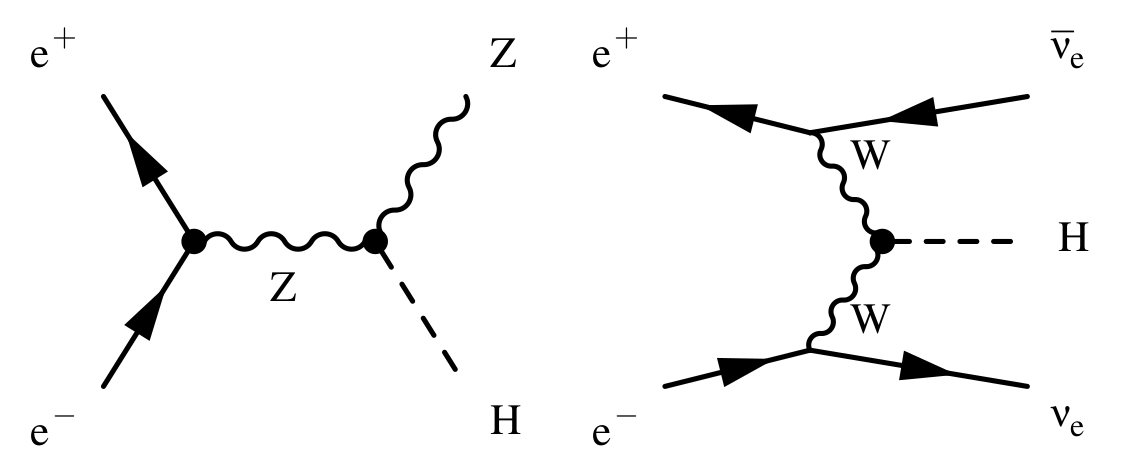}
\end{center}
\vspace*{-0.5cm}
  \caption{Leading-order Feynman diagrams of the highest
cross section Higgs production processes at CLIC: Higgsstrahlung
(left) and  WW-fusion (right).} 
  \label{fig:hdiag}
\end{figure}

Two processes dominate the Higgs boson production at \epem collisions,
see Fig.~\ref{fig:hdiag}.
For centre-of-mass energies up to about 450\,GeV, the Higgs-strahlung
process, Higgs boson production together with Z boson, dominates, see
Fig.~\ref{fig:htcs}.
%
%
Measurement of the Higgs boson production in the WW-fusion
process becomes increasingly important at higher energy stages,
however already at the initial ILC and CLIC stages it is helpful for
model-independent extraction of the Higgs boson couplings.

\subsection{Event reconstruction}

\begin{figure}[t]
\begin{center}
\includegraphics[height=5cm]{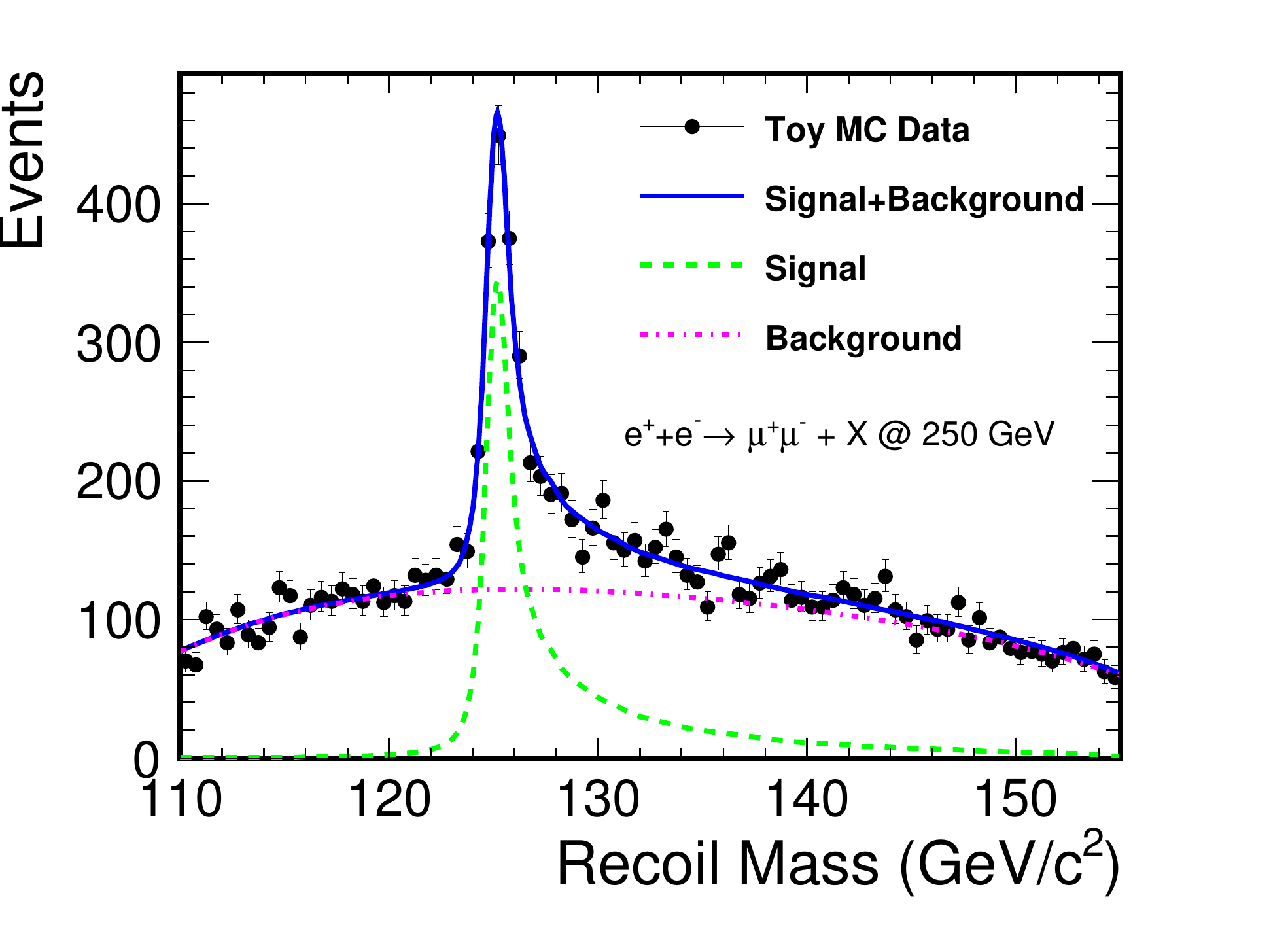}
\end{center}
\vspace*{-0.5cm}
  \caption{The recoil mass spectrum against $\PZ \to \PGmp\PGmm$ for
    $\PZ\PH$ signal and SM background, for ILC running at $\sqrt{s} =
    250$\,GeV \cite{Yan:2016xyx}. }
  \label{fig:mrec}
\end{figure}

\begin{figure}[p]
\begin{center}
 \includegraphics[width=0.3\textwidth]{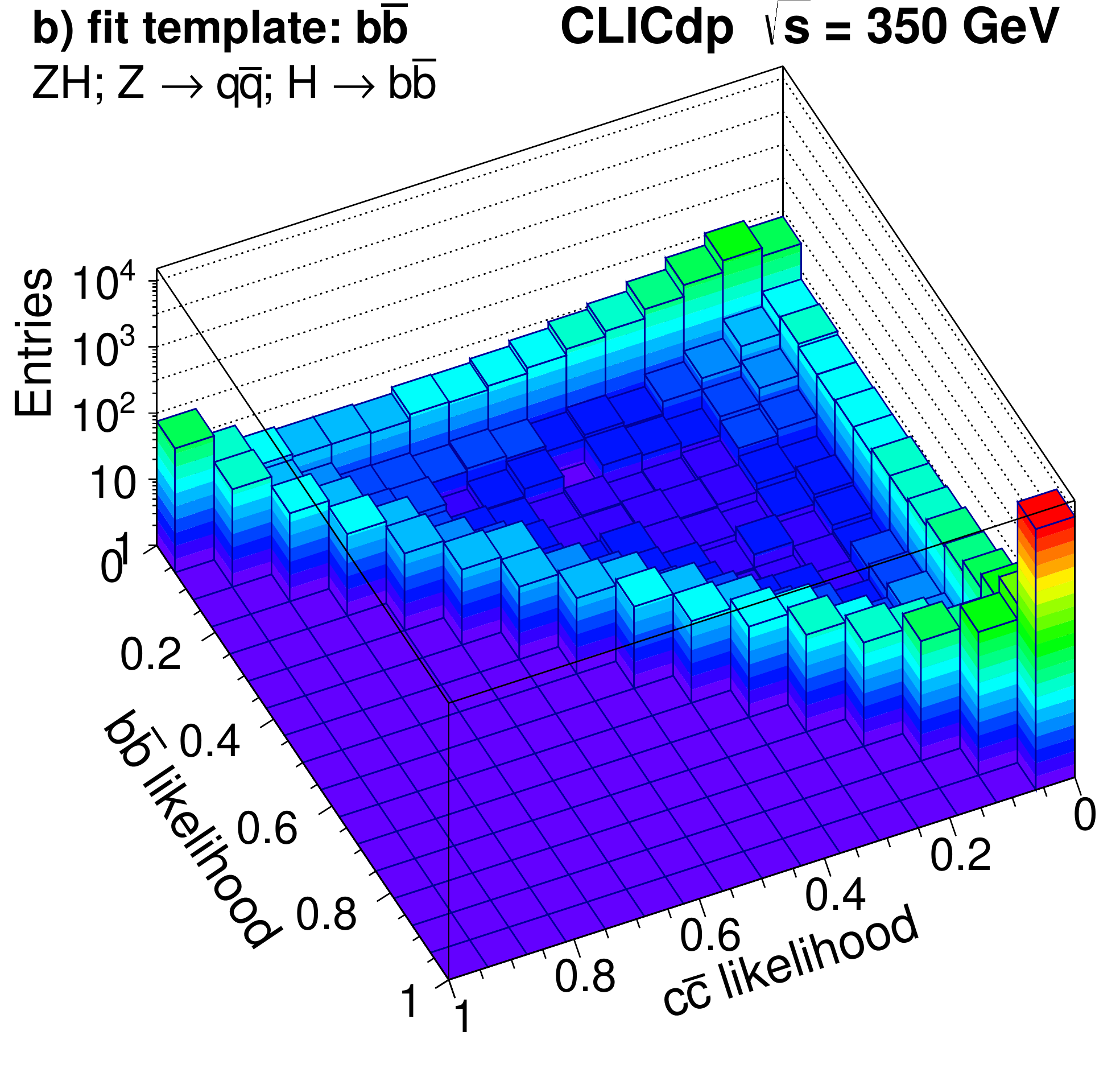} 
 \includegraphics[width=0.3\textwidth]{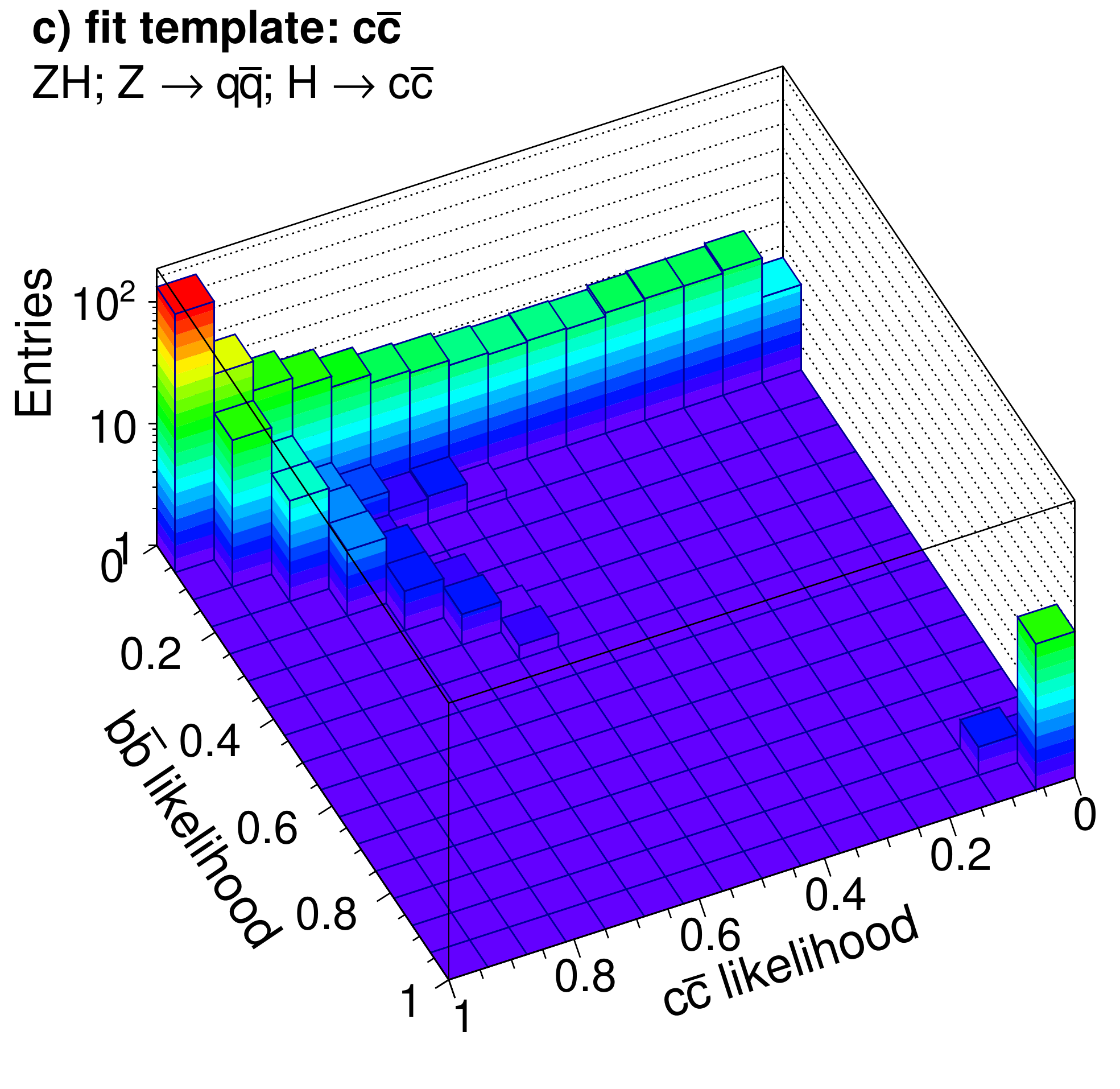} 
 \includegraphics[width=0.3\textwidth]{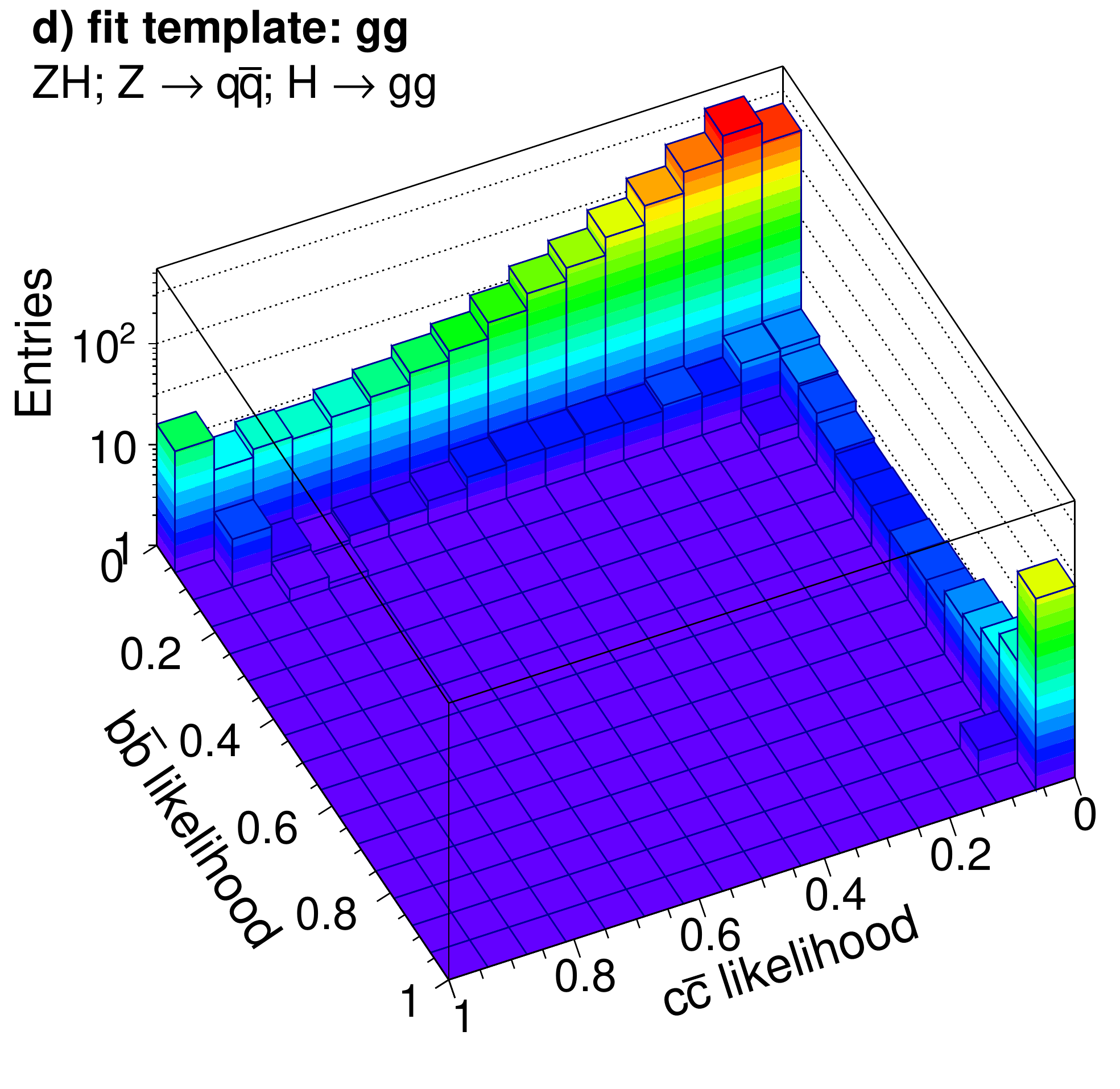} 
\end{center}
\vspace*{-0.5cm}
  \caption{Performance of the flavour tagging in Higgs boson decays:
$\PQb\PAQb$ likelihood versus $\PQc\PAQc$ likelihood distributions for
    $\Pep\Pem \to \PZ\PH$ events at $\sqrt{s} = 350$\,GeV, for the
    different event classes: $\PH \to \PQb\PAQb$ (left), $\PH \to
    \PQc\PAQc$ (center) and  $\PH \to \Pg\Pg$ (right) \cite{1608.07538}.
}
  \label{fig:htag}
\end{figure}

In the ZH production channel we can use ``Z-tagging'' approach for
unbiased selection of Higgs production events, i.e. making no
assumptions about the Higgs boson decays. 
Figure~\ref{fig:mrec} shows the recoil mass distribution for
Higgsstrahlung events and SM background processes, for a Z boson
decaying into a muon pair at the initial ILC stage \cite{Yan:2016xyx}.
We observe a sharp peak in the recoil mass distribution, which
allows model independent tagging of the Higgs boson production
events.
In the clean linear collider environment, 
unambiguous separation of different decay channels is then possible.
In particular, Higgs boson decays into two hadronic jets can be easily
separated into $\PQb\PAQb$, $\PQc\PAQc$ and $\Pg\Pg$ decays, as shown
in Fig.~\ref{fig:htag} \cite{1608.07538}.
This demonstrates clear prospects for direct measurement of
BR($\PH \to \PQc \PAQc$) and  BR($\PH \to \Pg \Pg$) at both ILC and CLIC.

  \subsection{Higgs couplings}

\begin{figure}[p]
\begin{center}
  \includegraphics[height=5cm]{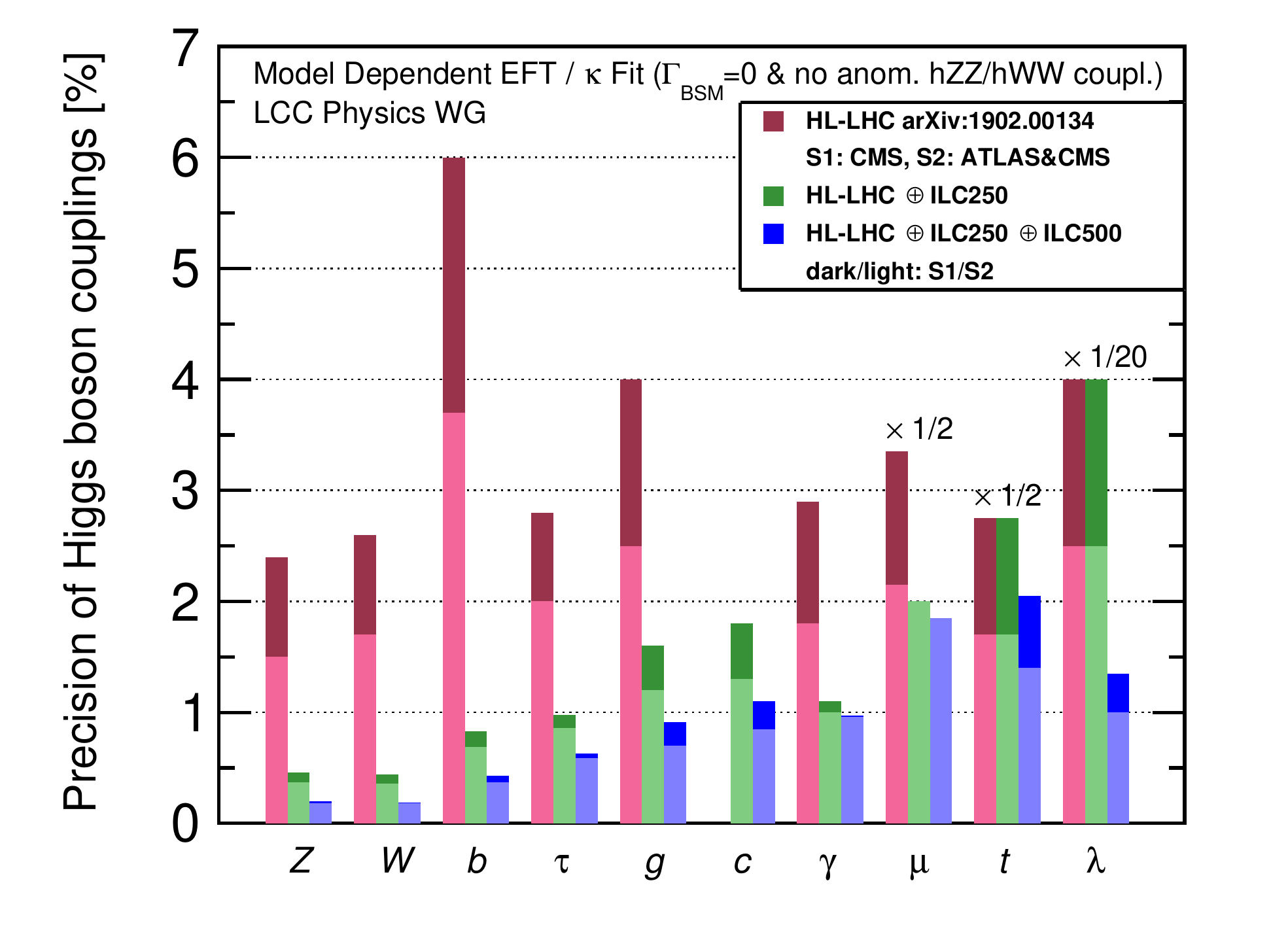}
  \includegraphics[height=4.8cm,width=0.46\textwidth]{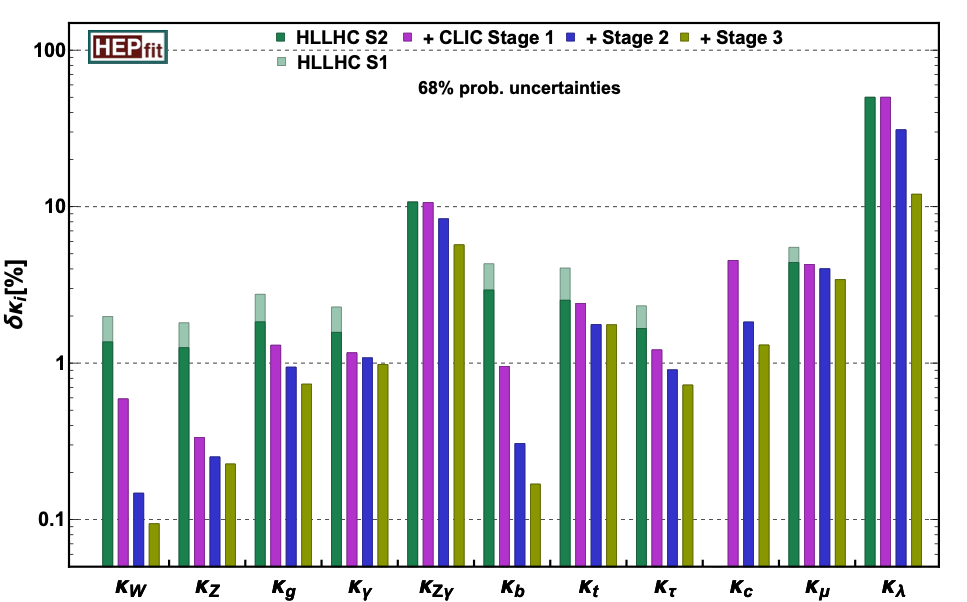}
\end{center}
\vspace*{-0.5cm}
  \caption{Precision of the Higgs couplings determined in a
    model-dependent fit, estimated for ILC \cite{1903.01629} (left)
    and CLIC \cite{1812.02093} (right) running at subsequent energy
    stages, compared with the expectations for the HL-LHC. 
  }
  \label{fig:hdep}
\end{figure}

\begin{figure}[p]
  \begin{center}
  \includegraphics[height=5cm]{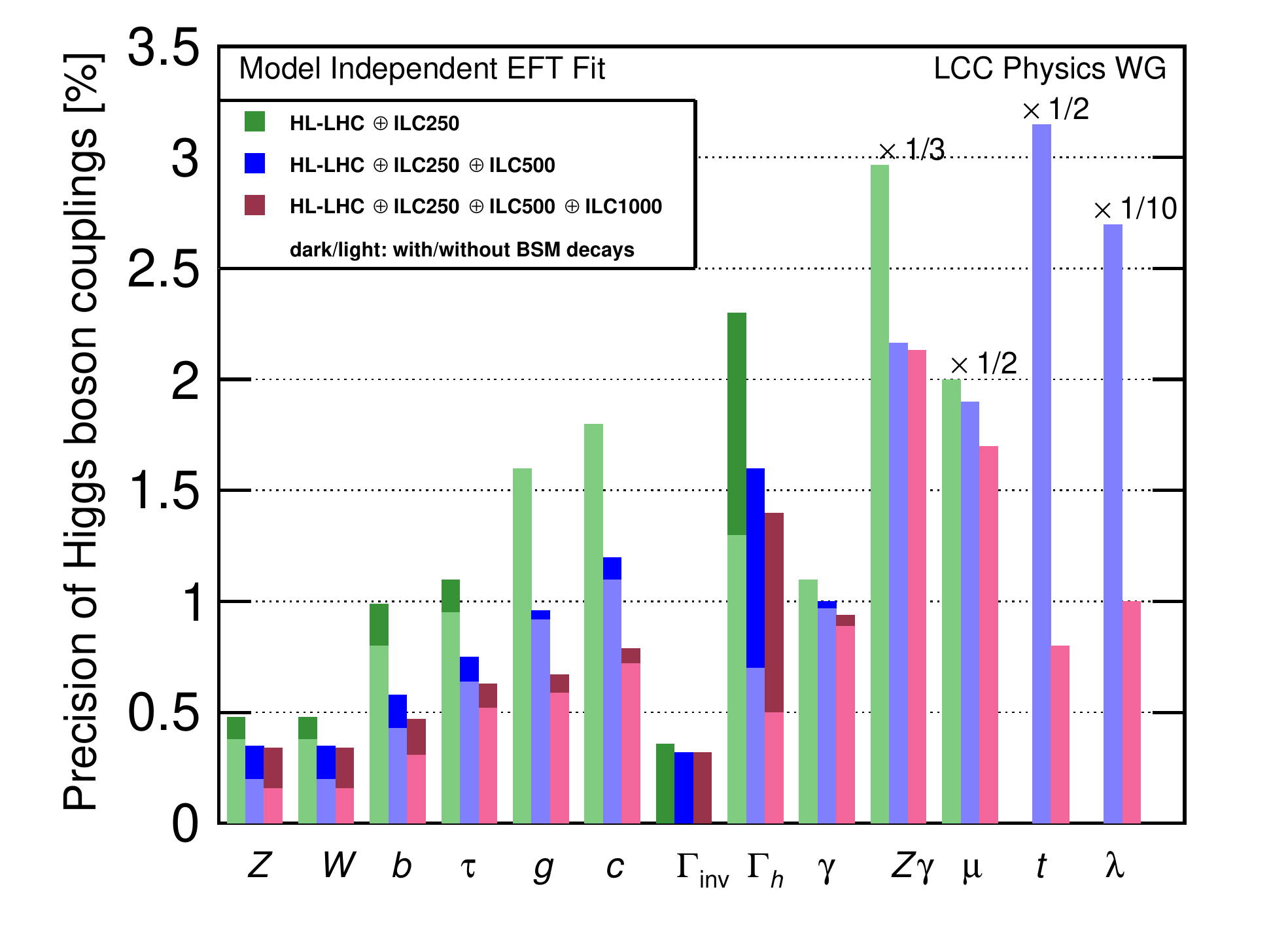}
  \includegraphics[height=5cm, trim = 8cm 0 0 0,clip ]{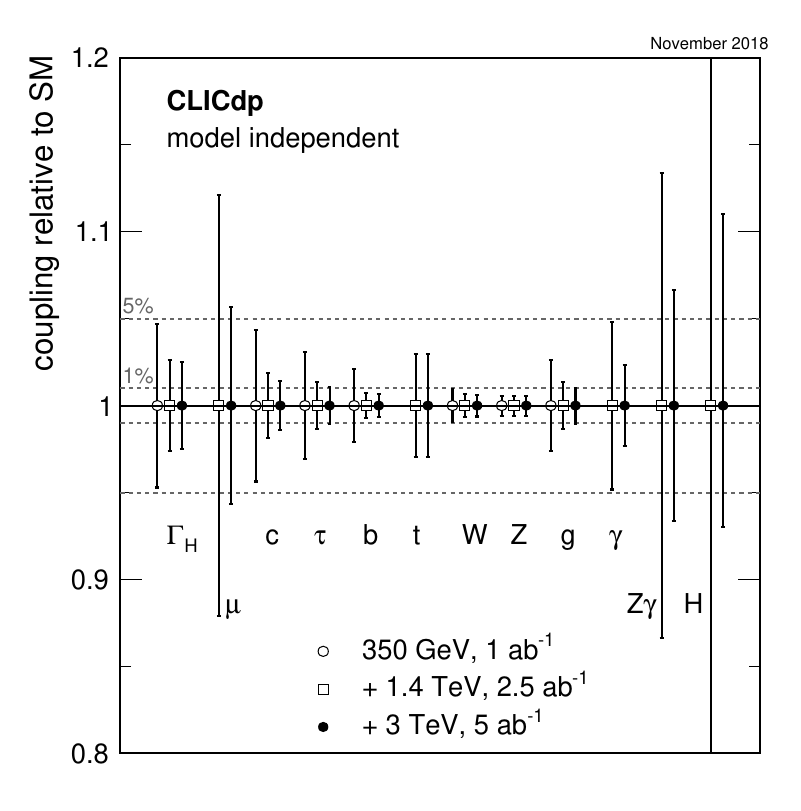} 
\end{center}
\vspace*{-0.5cm}
  \caption{Precision of the model-independent Higgs coupling
    determination from the SMEFT analysis at the ILC \cite{1903.01629} (left) 
    and from the $\kappa$-framework fit at CLIC \cite{1812.01644}
    (right), for subsequent energy stages.  
  }
  \label{fig:hind}
\end{figure}

  Future linear colliders guarantee sub-percent level precision for
  Higgs coupling measurements even at the first energy stages.
  The precision of Higgs boson coupling measurements at subsequent ILC 
  \cite{1903.01629} and  CLIC \cite{1812.02093,1812.01644} stages is summarised
  in Figs.~\ref{fig:hdep} and \ref{fig:hind}, for model-dependent and
  model-independent analysis approaches, respectively. 
  For the model-dependent coupling fit, where it is assumed
  that there are no non-Standard-Model Higgs decays nor anomalous
  couplings, the expected precision is up to an order of magnitude
  better than expected at HL-LHC.
  %
  %
  Two approaches to the model-independent coupling determination are
  possible:
  scaling factors for the SM Higgs boson couplings and its total width
  can be fitted to the Higgs boson measurements (so called
  $\kappa$-framework) \cite{1812.01644} or the global fit of all
  precision electroweak measurements can be performed in the framework
  of the Standard Model effective field theory (SMEFT)
  \cite{1708.08912,1903.01629}.
  Both approaches result in 
  increased coupling uncertainties, compared to model-dependent fit, but
  can also be used to constrain a wider range of BSM scenarios.
  
\begin{figure}[p]
\begin{center}
\includegraphics[height=5cm]{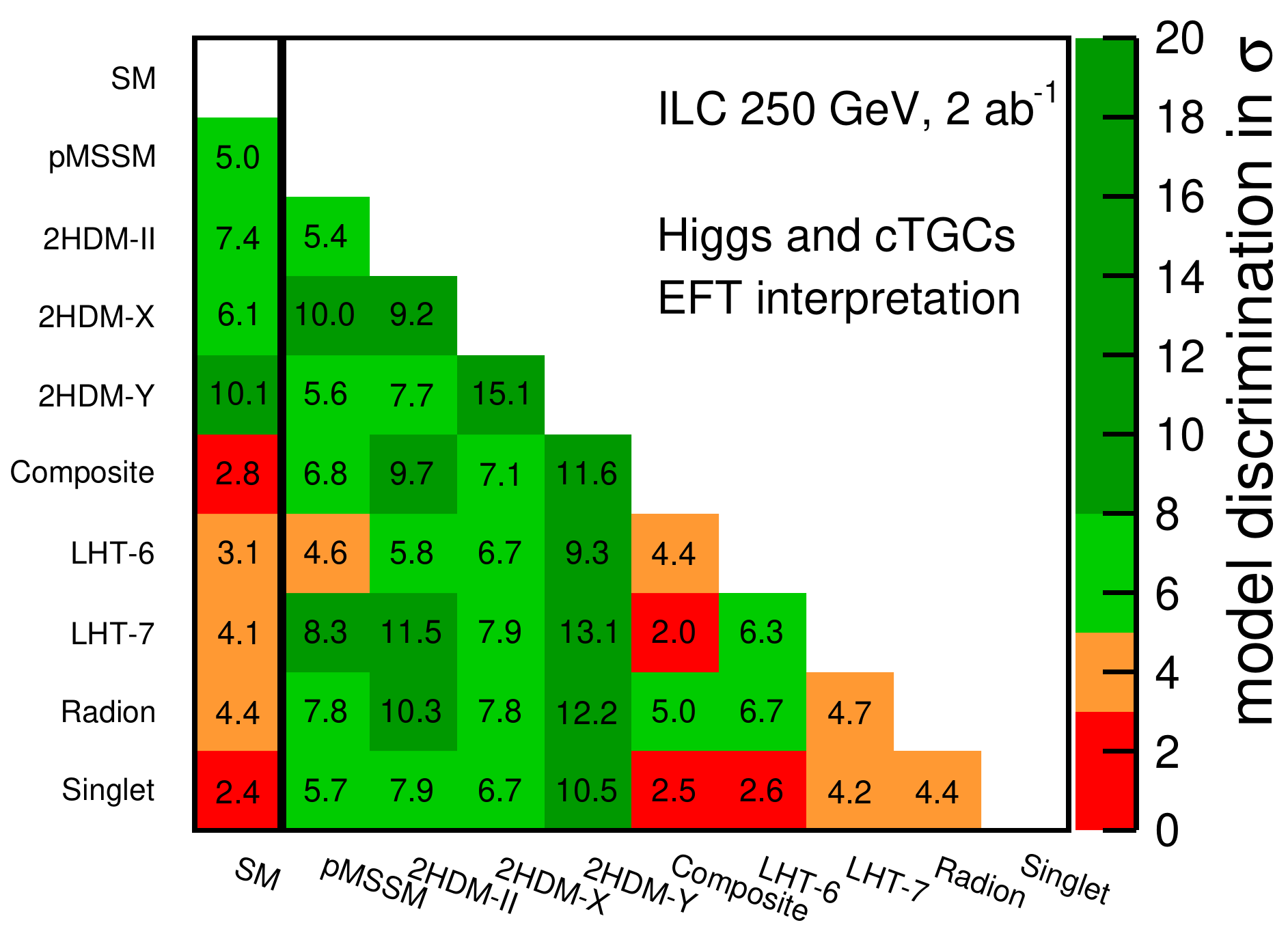}
\includegraphics[height=5cm]{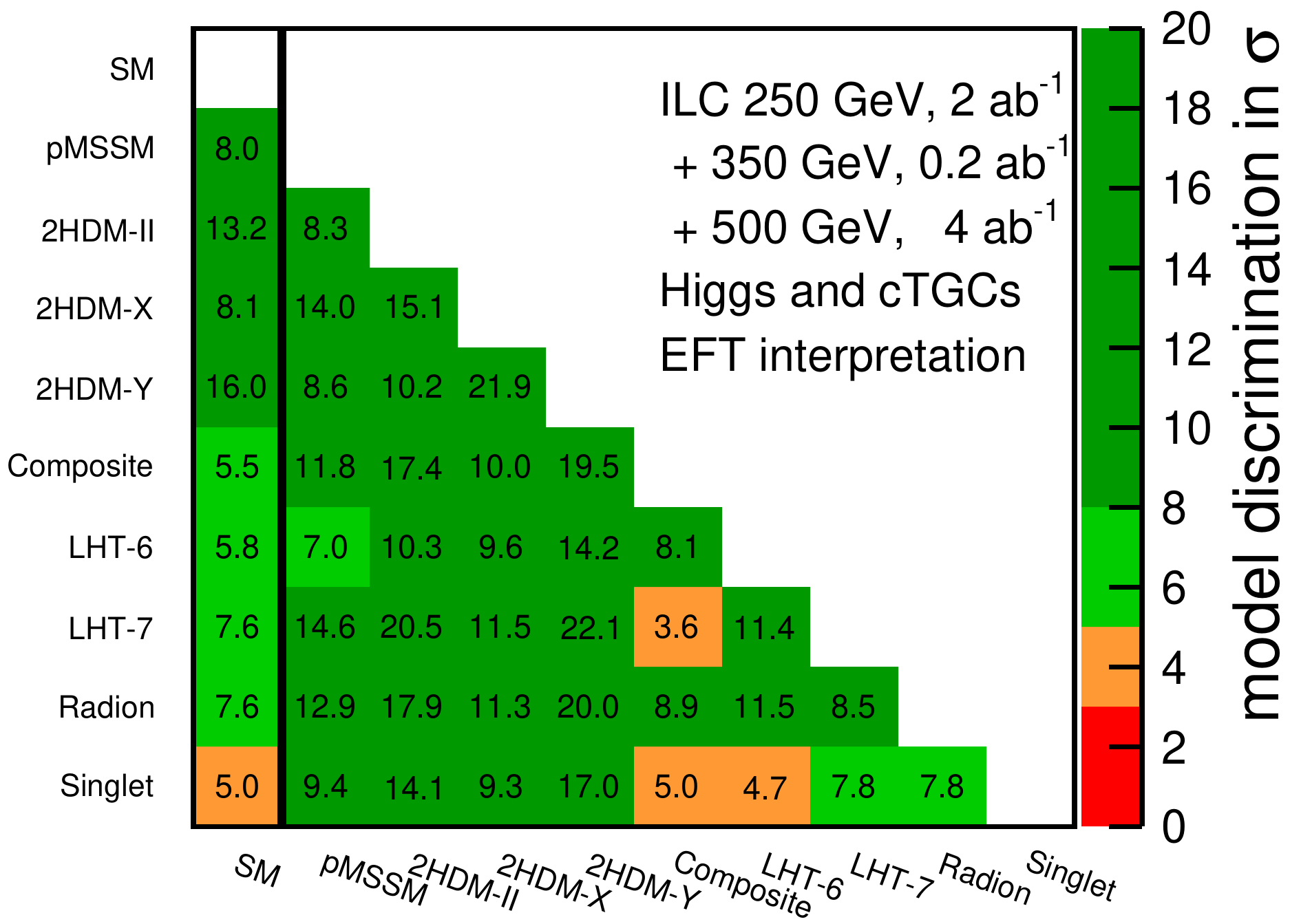}
\end{center}
\vspace*{-0.5cm}
  \caption{Expected discrimination power of the Higgs boson coupling
    fit at the ILC for the Standard Model and different BSM scenarios: 
 (left) with 2\,ab$^{-1}$ of data at 250\,GeV and (right)
with 2\,ab$^{-1}$ of data at 250\,GeV plus 4\,ab$^{-1}$ at 500\,GeV
\cite{1710.07621}.} 
  \label{fig:ilcbsm}
\end{figure}

The Standard Model gives exact predictions on the Higgs boson couplings to
all other SM particles.
Deviations from these predictions are expected in most BSM scenarios,
in particular in those with extended Higgs sector.
The precision of the linear \epem colliders should allow 
discrimination between the SM expectations and other models of ``new 
physics'' from the global analysis of the Higgs boson couplings. 
This is illustrated in Fig.~\ref{fig:ilcbsm} for the ILC running at
250\,GeV and for the combined 250\,GeV and 500\,GeV data \cite{1710.07621}.
Considered is a set of benchmark BSM models which are expected to escape
the direct search at HL-LHC.
Significant (above $5 \sigma$) discrimination between most scenarios 
will already be possible at 250\;GeV ILC.
After the full ILC programme, all BSM scenarios considered in the
study can be identified at $\ge 5 \sigma$ level.

\subsection{Invisible decays}

\begin{figure}[p]
\begin{center}
\includegraphics[height=5cm]{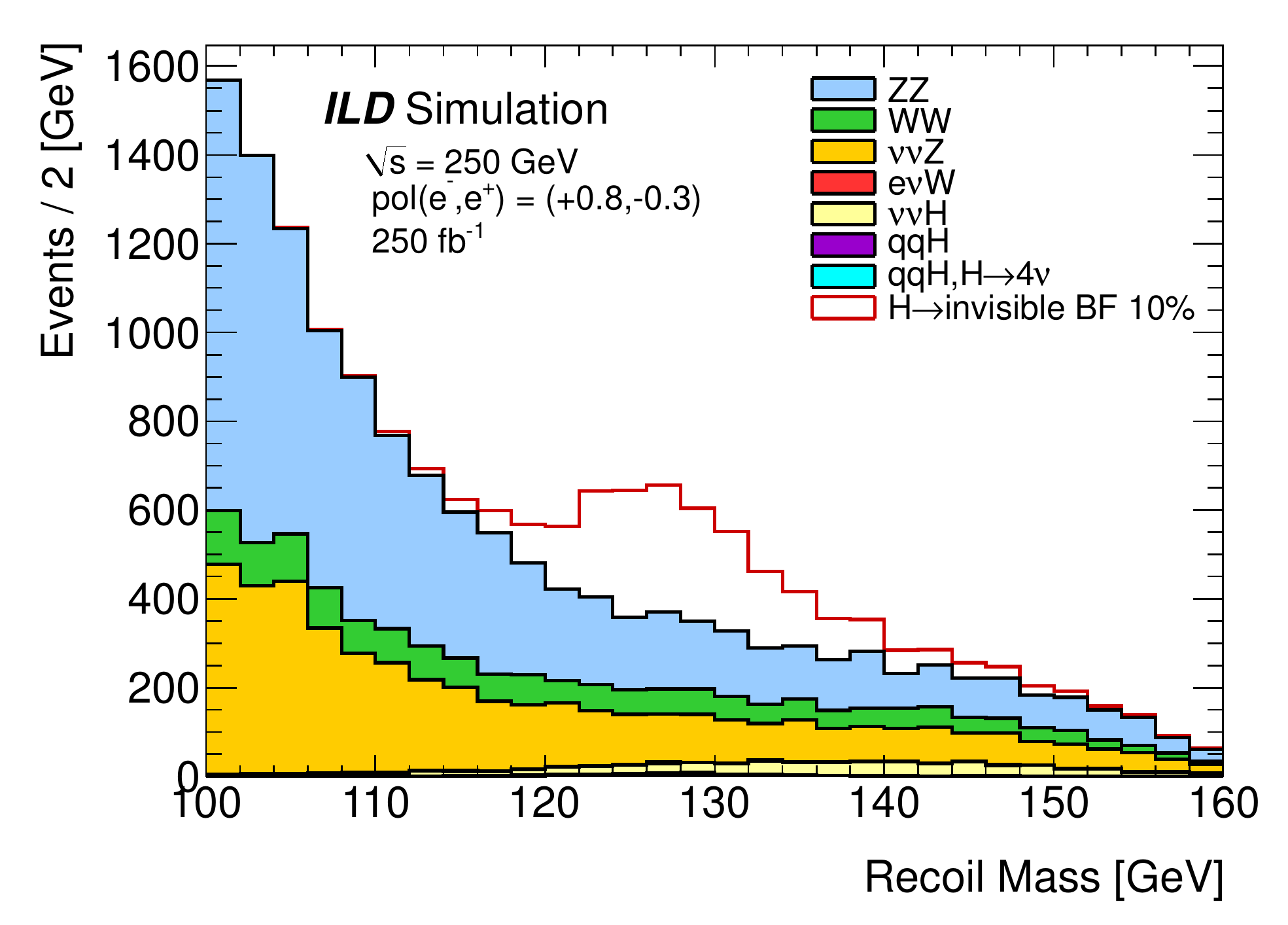}
\end{center}
\vspace*{-0.5cm}
  \caption{The recoil mass distribution for events with two jets in the
    final state, for ILC running at $\sqrt{s} = 250$\,GeV and
    right-handed beam polarisation \cite{1903.01629}. Open red
    histogram indicates the expected contribution from invisible Higgs
  boson decays for BR($\PH \to$ inv) = 10\%.}
  \label{fig:hinv}
\end{figure}

As already mentioned above, Higgs boson production in the
Higgsstrahlung process allows  a model independent determination
of the Higgs boson properties at future linear colliders. 
Large samples of events can be selected when the hadronic decay channel of
the $Z$ boson is considered. 
Events with mono-$Z$ production, and no other activity in the
detector, can be considered as candidate events for the invisible
Higgs boson decays, if the recoil mass reconstructed from
energy-momentum conservation is consistent with the Higgs boson 
mass. 
Figure~\ref{fig:hinv} shows the expected recoil mass distribution
for events with two jets in the final state, for ILC running at
$\sqrt{s} = 250$\,GeV \cite{1903.01629}.
The main SM background processes in this analysis are the production
of $\PZ\PZ$ and $\PWp\PWm$ pairs, as well as single $\PZ$ production
via the $\PW\PW$ fusion, $\Pep\Pem \to \PGne\PAGne\PZ$.
Also indicated is the expected contribution of invisible Higgs boson
decays, assuming branching ratio BR($\PH \to$ inv) = 10\%.
With 2\,ab$^{-1}$ collected at 250\,GeV ILC the expected 95\%
C.L. limit on invisible decays of the 125\,GeV Higgs boson is 0.23\%
\cite{ilchinv2}.

\subsection{Higgs boson self-coupling measurement}

Measurement of the trilinear Higgs boson coupling, which provides a
direct probe of the shape of the Higgs potential, is a crucial test of
the Standard Model and of the electroweak symmetry breaking mechanism
in general. 
The coupling can be constrained using indirect measurements, as the
triple Higgs coupling does affect single Higgs boson production via
radiative corrections. 
However,  more robust determination of the triple Higgs coupling is
possible by direct measurement of Higgs pair production processes
involving the trilinear coupling at the tree level.

\begin{figure}[p]
\begin{center}
\includegraphics[height=5cm]{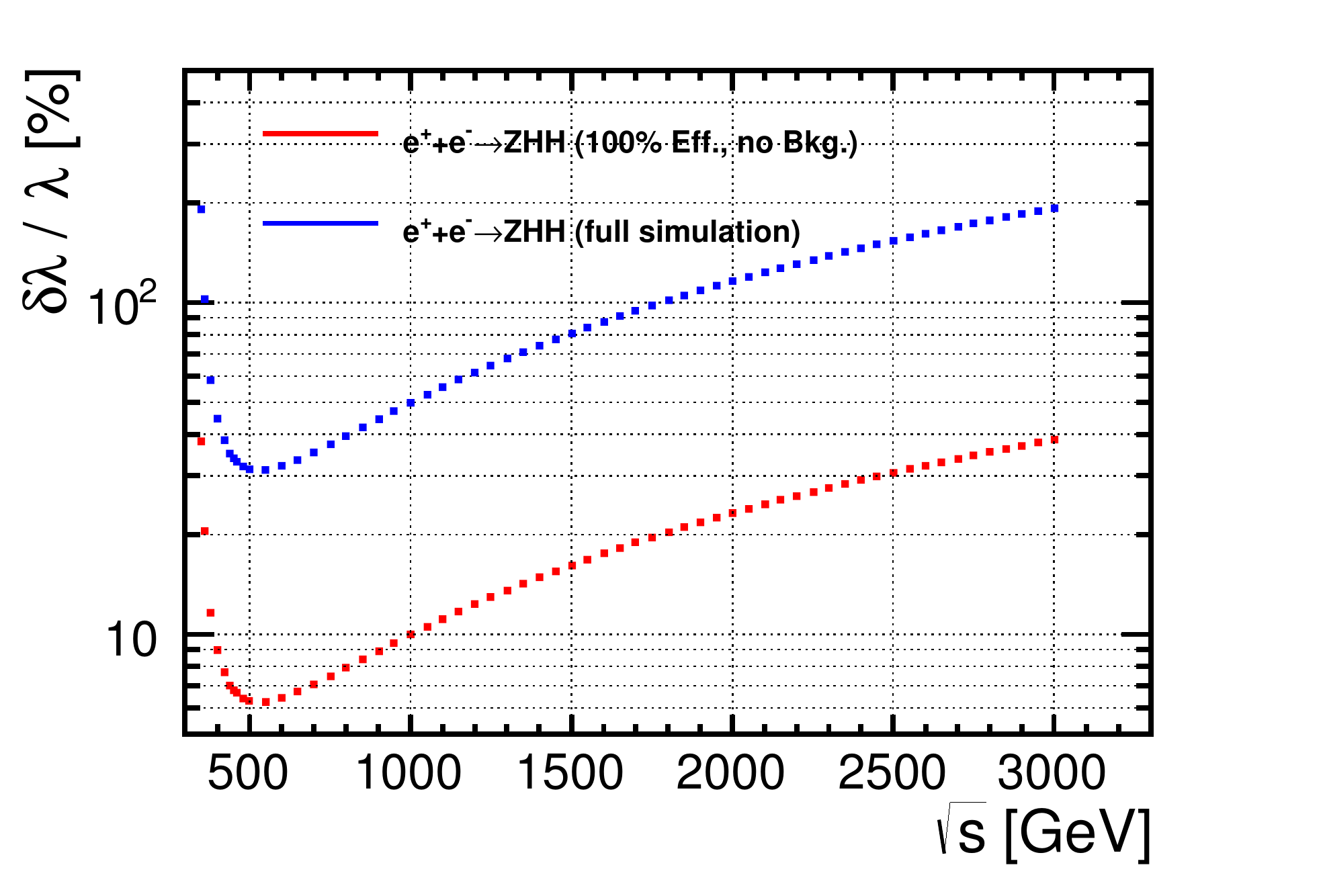}
\includegraphics[height=5cm]{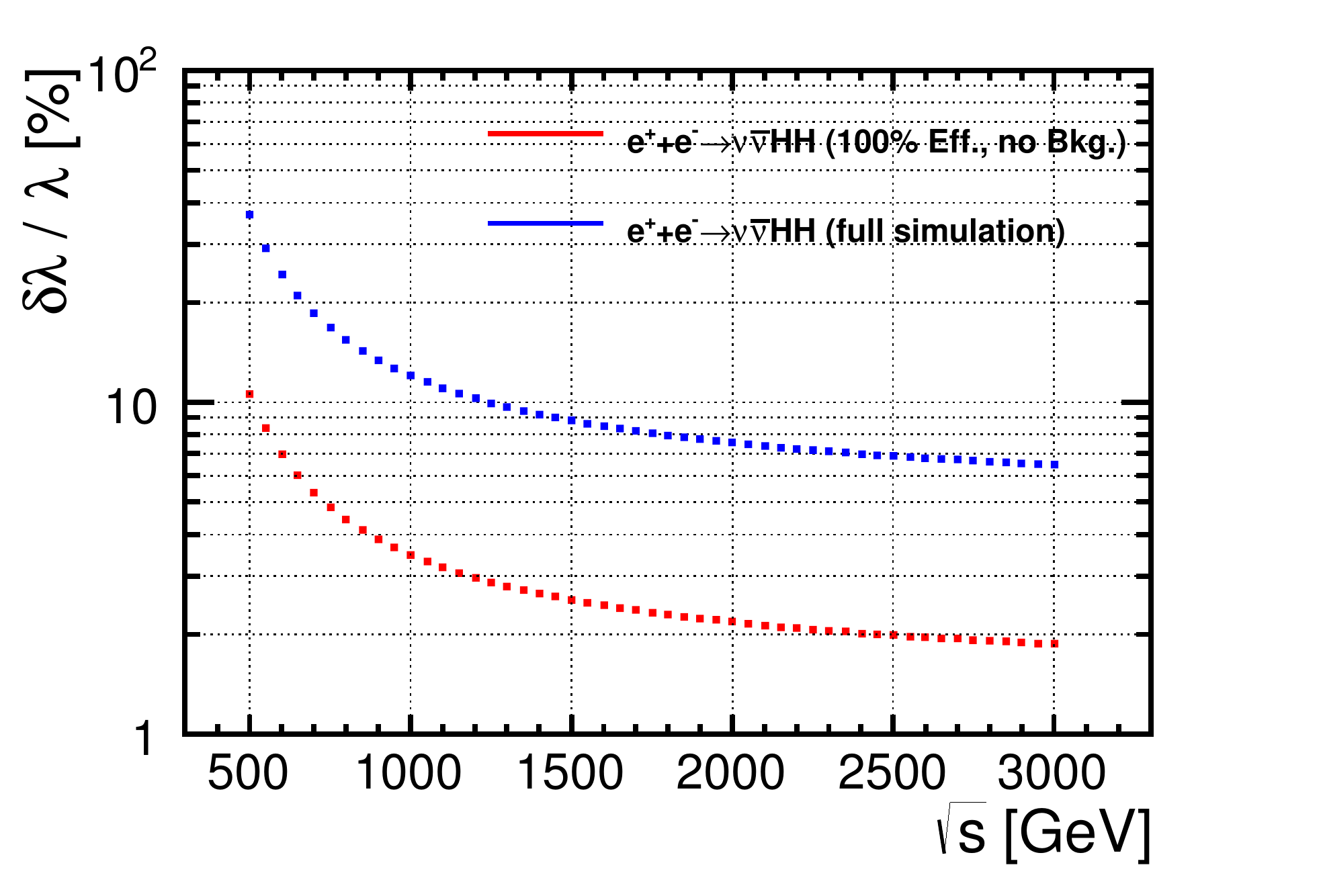}
\end{center}
\vspace*{-0.5cm}
  \caption{Estimated precision on the determination of the Higgs
    self-coupling $\lambda$ from the measurement of the double Higgs
    production: $\Pep\Pem\to\PZ\PH\PH$ (left) and
    $\Pep\Pem\to\PGn\PAGn\PH\PH$ (right), as a function of the
    centre-of-mass energy. Same integrated luminosities of
    4\,ab$^{-1}$ is assumed at all $\sqrt{s}$ \cite{1903.01629}. }
  \label{fig:dihprod}
\end{figure}
\begin{figure}[p]
\begin{center}
  \includegraphics[height=5cm]{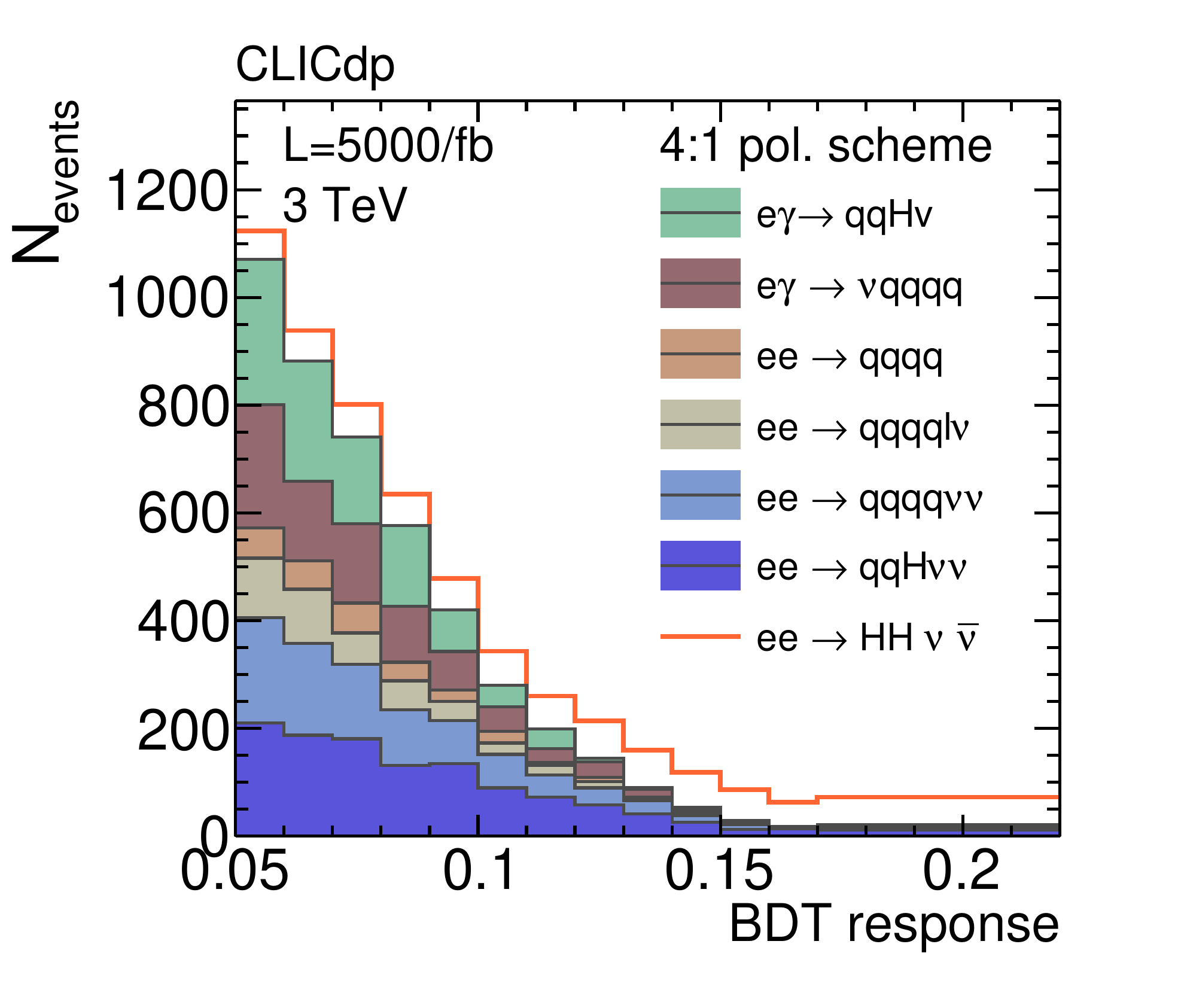}
  \includegraphics[height=5cm]{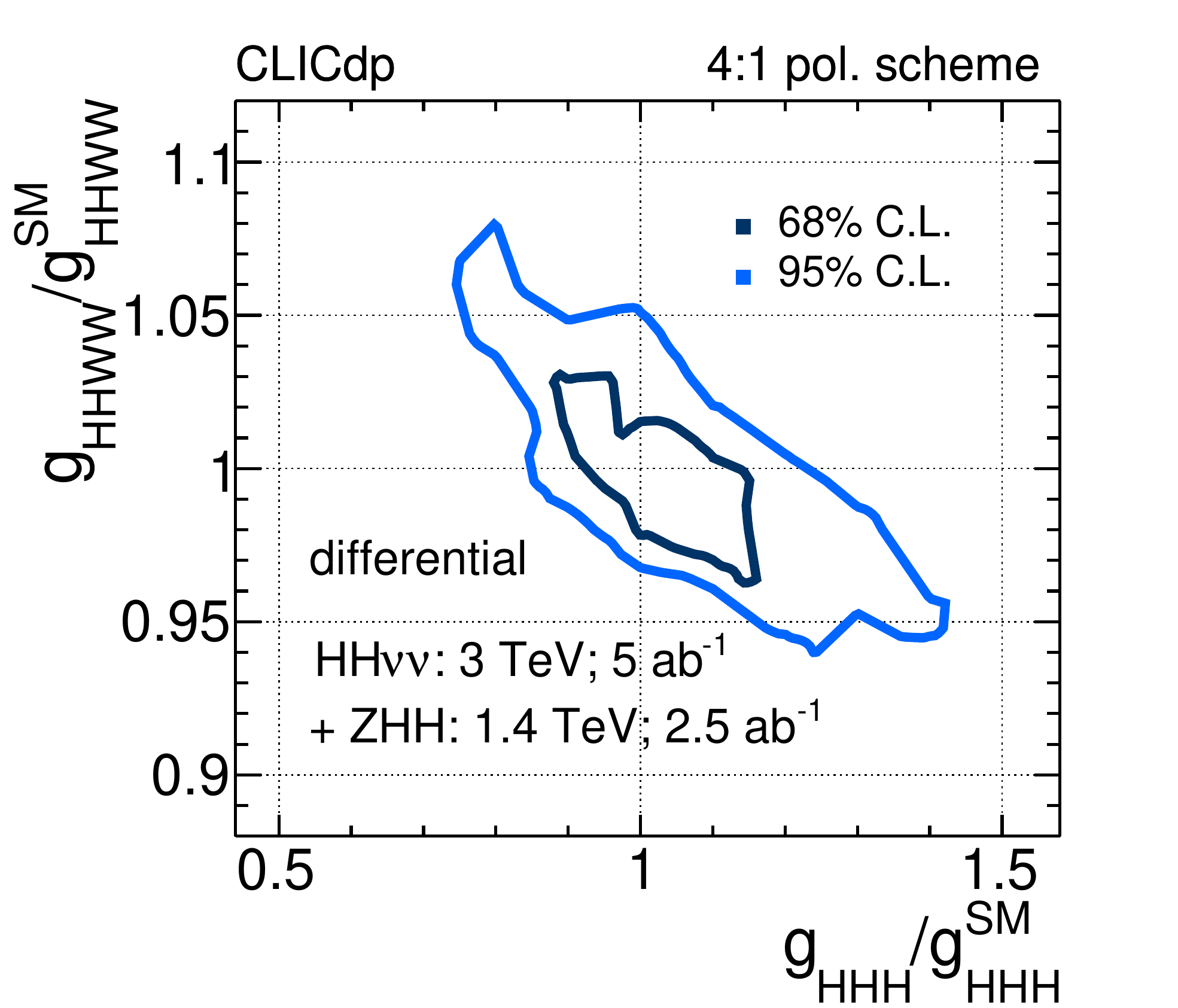}
\end{center}
\vspace*{-0.5cm}
  \caption{Left: BDT response distribution of all SM contributions
    stacked in the loose BDT selection at 3\,TeV CLIC. Right:
    confidence contours at 68\% and 95\% C.L. for the simultaneous fit
    of $\kappa_{HHH}$ and $\kappa_{HHWW}$ based on CLIC measurements
    at 1.4\,TeV and 3\,TeV \cite{1901.05897}.} 
  \label{fig:trih}
\end{figure}

Two processes are considered for direct Higgs self-coupling
measurement:
double Higgs boson production in the Higgsstrahlung-like process,
$\Pep\Pem \to \PZ \PH \PH$ and
double Higgs boson production in the WW-fusion process,
$\Pep\Pem \to \PGn \PAGn \PH\PH$, which becomes important only at 
high beam energies, see Fig.~\ref{fig:htcs}.
Figure~\ref{fig:dihprod} presents the expected uncertainties in
the Higgs self-coupling determination for these two processes, as a
function of the centre-of-mass energy \cite{1903.01629}.
ILC running at 500\,GeV seems to be optimal for the measurement of the
trilinear Higgs boson coupling in the $\Pep\Pem \to \PZ \PH \PH$
channel. 
The dedicated study indicates that the total cross section for this
process can be measured with 4\,ab$^{-1}$ to about 17\% corresponding
to 27\% uncertainty on the trilinear coupling value \cite{1903.01629}.
For higher centre-of-mass energies, better accuracy can be expected
from the measurement of $\Pep\Pem\to\PGn\PAGn\PH\PH$.
Prospects for constraining both the trilinear Higgs self-coupling and
the quartic $\PH\PH\PW\PW$ coupling from the measurement of double
Higgs boson production at CLIC were studied in \cite{1901.05897}.
Figure~\ref{fig:trih} gives the BDT response distribution expected
at 3\,TeV CLIC and the confidence contours at 68\% and 95\% C.L. for
the simultaneous fit of the trilinear coupling, $\kappa_{HHH}$, and
quartic coupling, $\kappa_{HHWW}$, based on CLIC measurements at
1.4\,TeV and 3\,TeV. 
The trilinear Higgs self-coupling can be constrained to
 $\delta\lambda / \lambda =  -7 \% / +11 \%$ (68\% C.L.) \cite{1901.05897}.

\subsection{Looking for BSM effects}

\begin{figure}[p]
\begin{center}
   \includegraphics[height=5cm]{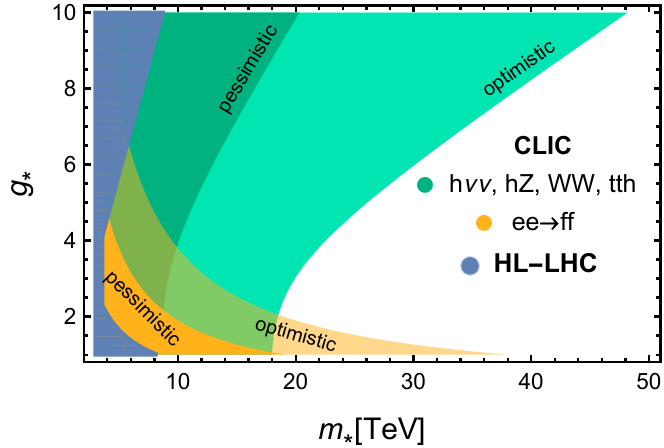}
\end{center}
\vspace*{-0.5cm}
  \caption{Expected 5$\sigma$ discovery contours for Higgs
    compositeness in the $(m^\star\!\!,\, g^\star )$ plane, overlaid with
    the 2$\sigma$ projected exclusions from HL-LHC  \cite{1812.07986}.} 
  \label{fig:comph}
\end{figure}

All precision measurements, including Higgs boson production and other
relevant SM processes,  can be combined in a more general analysis
based on the effective field theory (EFT) approach. 
An example of such an analysis for CLIC is presented in Fig.~\ref{fig:comph}
for the composite Higgs model \cite{1812.02093,1812.07986}.
The model is characterised by the typical composite-sector mass
$m^\star$ and coupling $g^\star$. 
The expected CLIC discovery range (5$\sigma$) for Higgs compositeness 
is compared to expected HL-LHC 2$\sigma$ exclusion limits.
CLIC running at TeV energies can be sensitive to the compositeness mass
scales up to about 50\,TeV going far beyond the HL-LHC reach for this
model.


\begin{figure}[p]
\begin{center}
\includegraphics[height=5cm]{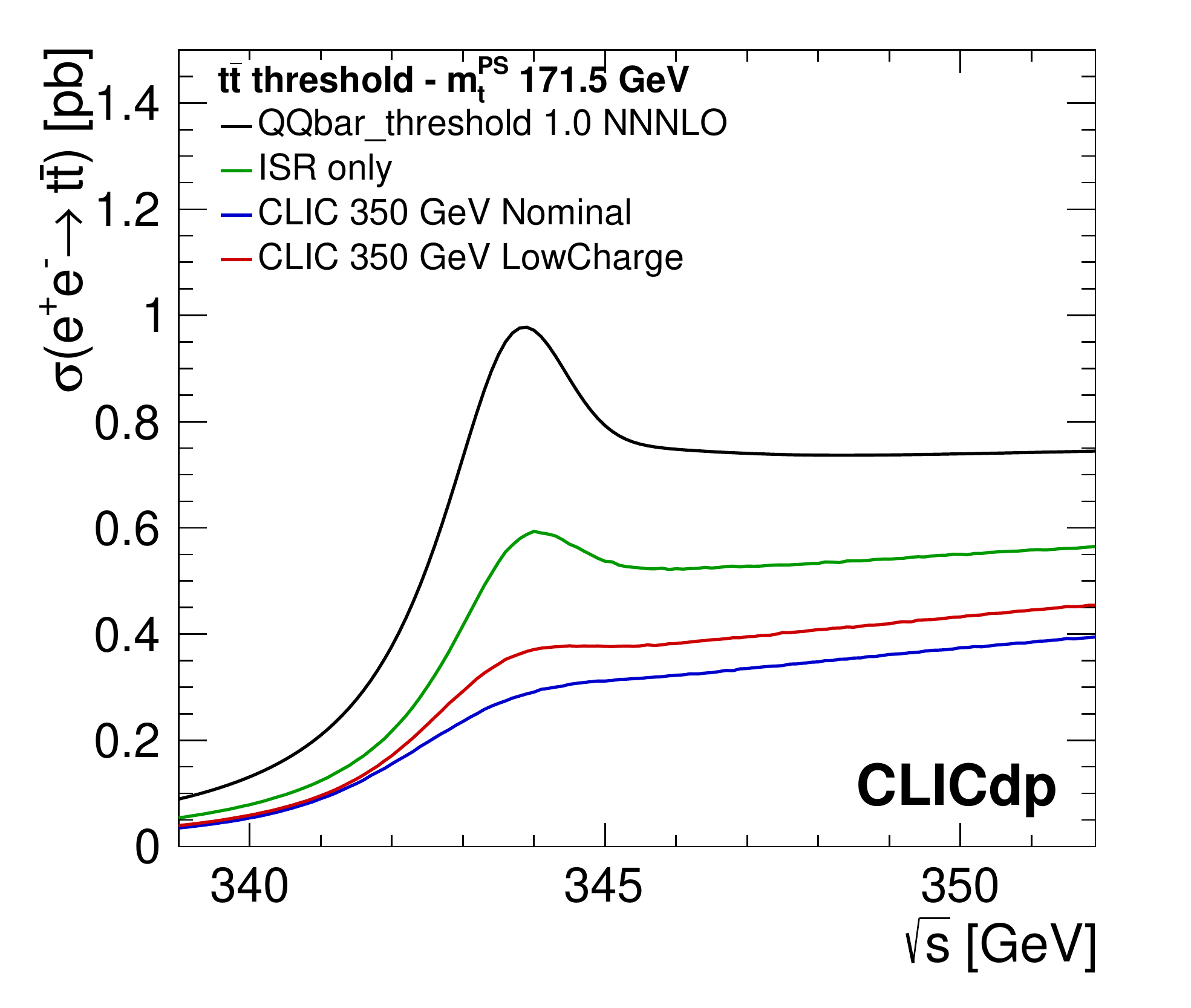}
\end{center}
\vspace*{-0.5cm}
  \caption{The impact of the initial state radiation (ISR) and the
    luminosity spectra on the top-quark pair production cross section
    at the threshold. The blue and red curves show the observable
    cross section for the nominal and the ``reduced charge'' 
   luminosity spectra for CLIC, respectively \cite{1807.02441}.} 
  \label{fig:thrth}
\end{figure}

\section{Top-quark physics}

With an expected value of the Yukawa coupling of the order of one, the
precise determination of the top quark properties is crucial for the
understanding of electroweak symmetry breaking and of the vacuum
stability of the Standard Model. 
As the top quark gives large loop contributions to many
precision measurements, determination of its properties is also
essential for many ``new physics'' searches.

\subsection{Top-quark mass}

The dependence of the theoretical top pair production cross section on the
centre-of-mass energy shows a clear resonance-like structure at the threshold,
corresponding to a narrow \ttbar  bound state.
The threshold cross section is strongly affected by the smearing due
to luminosity spectra (which can be partly reduced by using dedicated
running configuration) and initial state radiation (ISR), see
Fig.~\ref{fig:thrth} \cite{1807.02441}.  
The shape of the cross section is also very sensitive to other top quark
properties and model parameters: top-quark width $\Gamma_t$,
Yukawa coupling $y_t$ and strong coupling constant $\alpha_s$.
Still, the threshold scan is considered the best approach to determine
the top-quark mass, $m_t$, with highest possible precision. 
The other advantage the threshold scan is that the extracted top-quark
mass parameter is well defined from the theoretical point of view.

\begin{figure}[tb]
\begin{center}
\includegraphics[height=5cm]{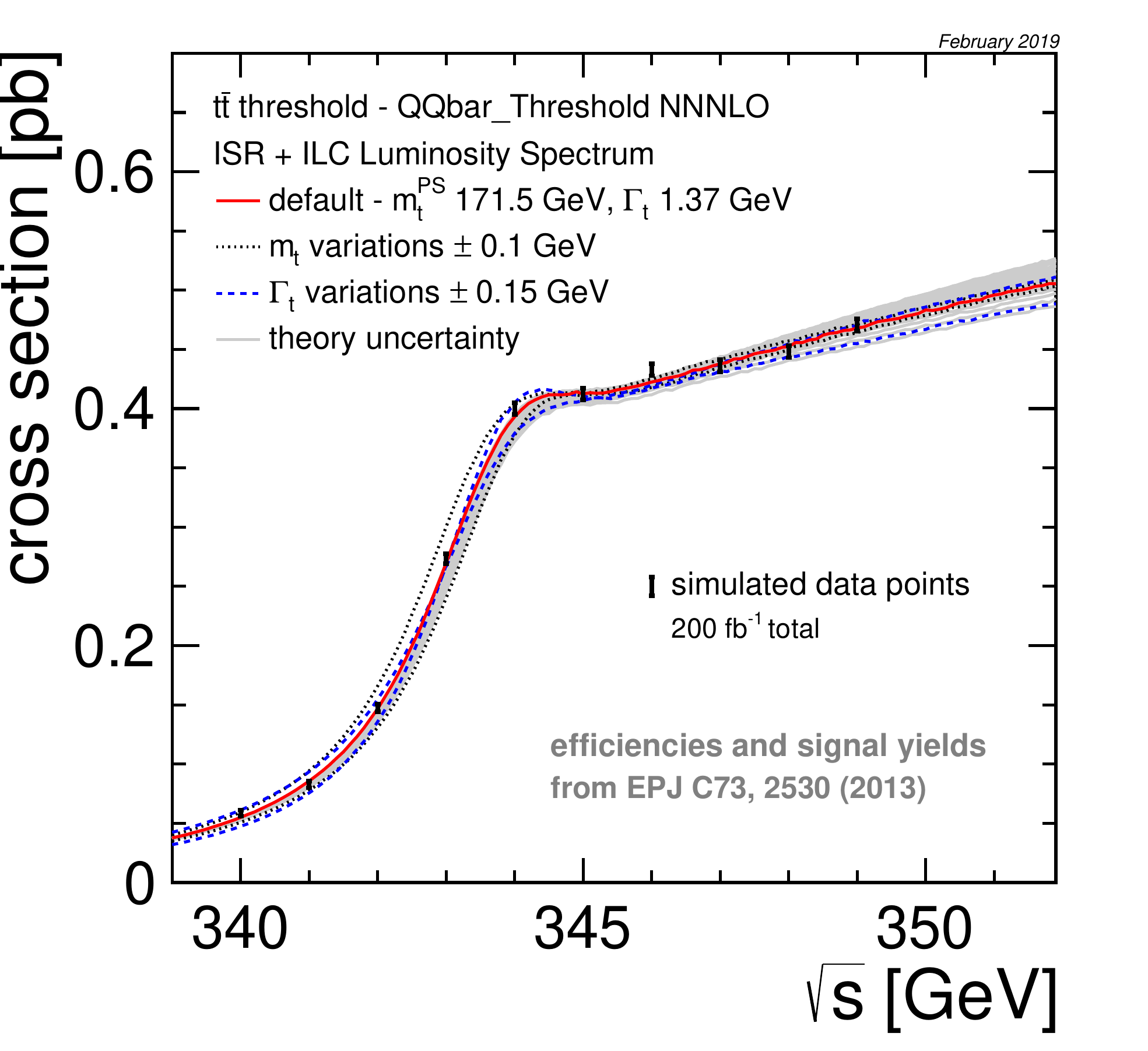}
\includegraphics[height=5cm]{plots/mapa_yuk_alfa_new.png}
\end{center}
\vspace*{-0.5cm}
  \caption{Left: reference scenario for the top-quark threshold scan
    considered for the ILC \cite{1903.01629}.
    Right: expected statistical uncertainty on
    the top-quark mass from the fit of four model parameters
    ($m_t$, $\Gamma_t$, $y_t$  and $\alpha_s$) as a function of the
    uncertainties on the strong coupling constant and the Yukawa
    coupling values from earlier measurements, assuming total
    normalisation uncertainty  (data + theory) of 0.1\%
    \cite{Wilga2019}.
  }  
  \label{fig:thrfit}
\end{figure}

Baseline threshold scan scenario assumes running at 10 equidistant
energy points taking 10-20\,fb$^{-1}$ of data for each value of
$\sqrt{s}$.
Such a scenario is indicated in Fig.~\ref{fig:thrfit} (left) \cite{1903.01629}.
A statistical uncertainty of about 20\,MeV is expected from a mass and
width fit and the total systematic uncertainty should be controlled
to the level of 50\,MeV \cite{1903.01629,1807.02441}.
However, to reach this level of statistical uncertainty, the strong
coupling constant and the top-quark Yukawa coupling need to be constrained
from independent measurements, see Fig.~\ref{fig:thrfit} (right) 
\cite{Wilga2019}.
Longitudinal beam polarisation has not been considered in these studies.

\begin{figure}[tb]
\begin{center}
\includegraphics[height=5cm, trim = 8cm 0 0 0 , clip]{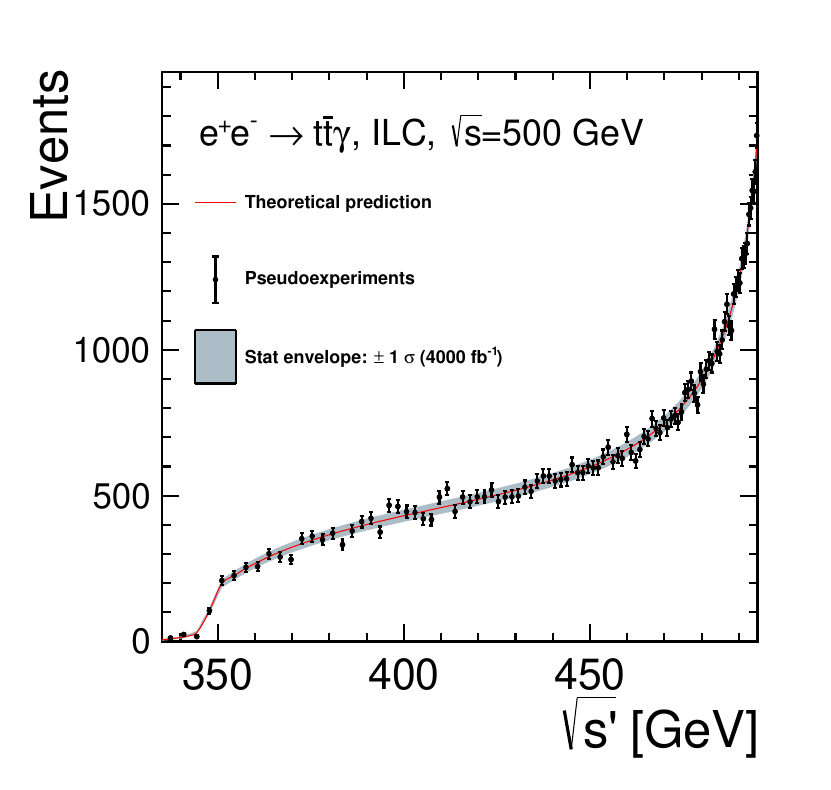}
\includegraphics[height=5cm]{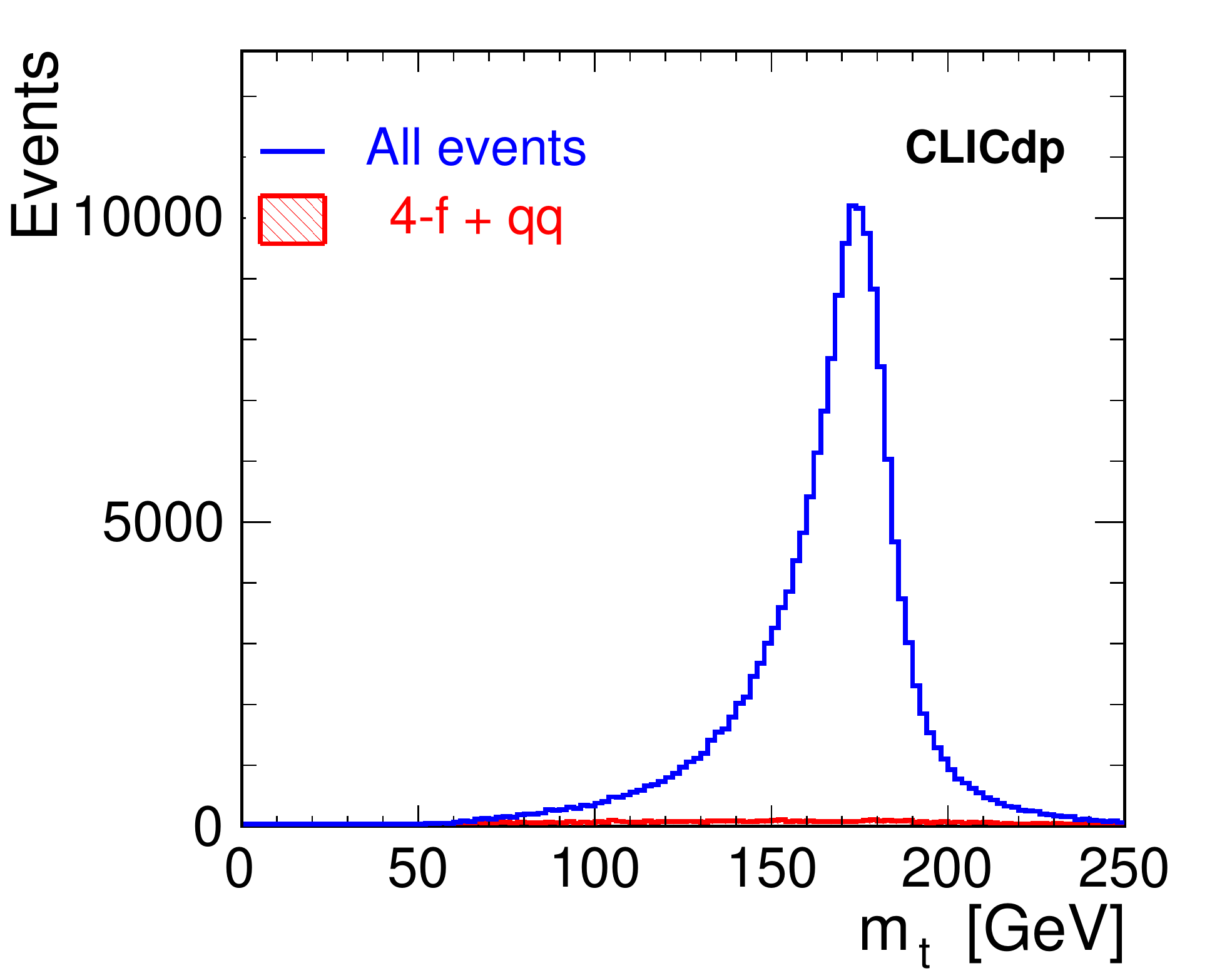}
\end{center}
\vspace*{-0.5cm}
  \caption{Expected distributions of (left) the effective \ttbar
    centre-of-mass energy for radiative events at 500\,GeV ILC
    \cite{1912.01275} and (right)  distribution of the top-quark mass
    reconstructed for the semi-leptonic  samples of top-quark
    pair-production events at 380\,GeV CLIC \cite{1807.02441}.}  
  \label{fig:mtop}
\end{figure}

The top-quark mass can also be extracted from measurements at higher
centre-of-mass energies.
One of the options is the threshold cross section determination
from radiative events $\Pep\Pem\to\PQt\PAQt\PGg$, see
Fig.~\ref{fig:mtop} (left).
Measurement of a threshold in the reconstructed \ttbar invariant mass
should allow for top-quark mass extraction with a statistical
precision of the order of 90\,MeV for CLIC running at 380\,GeV and
110\,MeV for ILC running at 500\,GeV, for integrated luminosities of
1\,ab$^{-1}$ and 4\,ab$^{-1}$, respectively \cite{1912.01275}. 
Taking systematic effects into account, the total uncertainty of the
extracted top-quark mass is  110\,MeV at the initial stage of CLIC and
150\,MeV for ILC running at 500\,GeV. 
The top-quark mass can be also measured directly from reconstruction of
hadronic top-quark decays, see Fig.~\ref{fig:mtop} (right). 
For the initial stage of CLIC, the statistical precision of about
30\,MeV is expected for combined hadronic and semi-leptonic samples of
top-quark pair-production events \cite{1807.02441}.
The measurement requires excellent control of the jet energy
scale and the extracted mass is also subject to large theoretical
uncertainties.


\subsection{FCNC top-quark decays}

\begin{figure}[tb]
\begin{center}
\includegraphics[height=5cm]{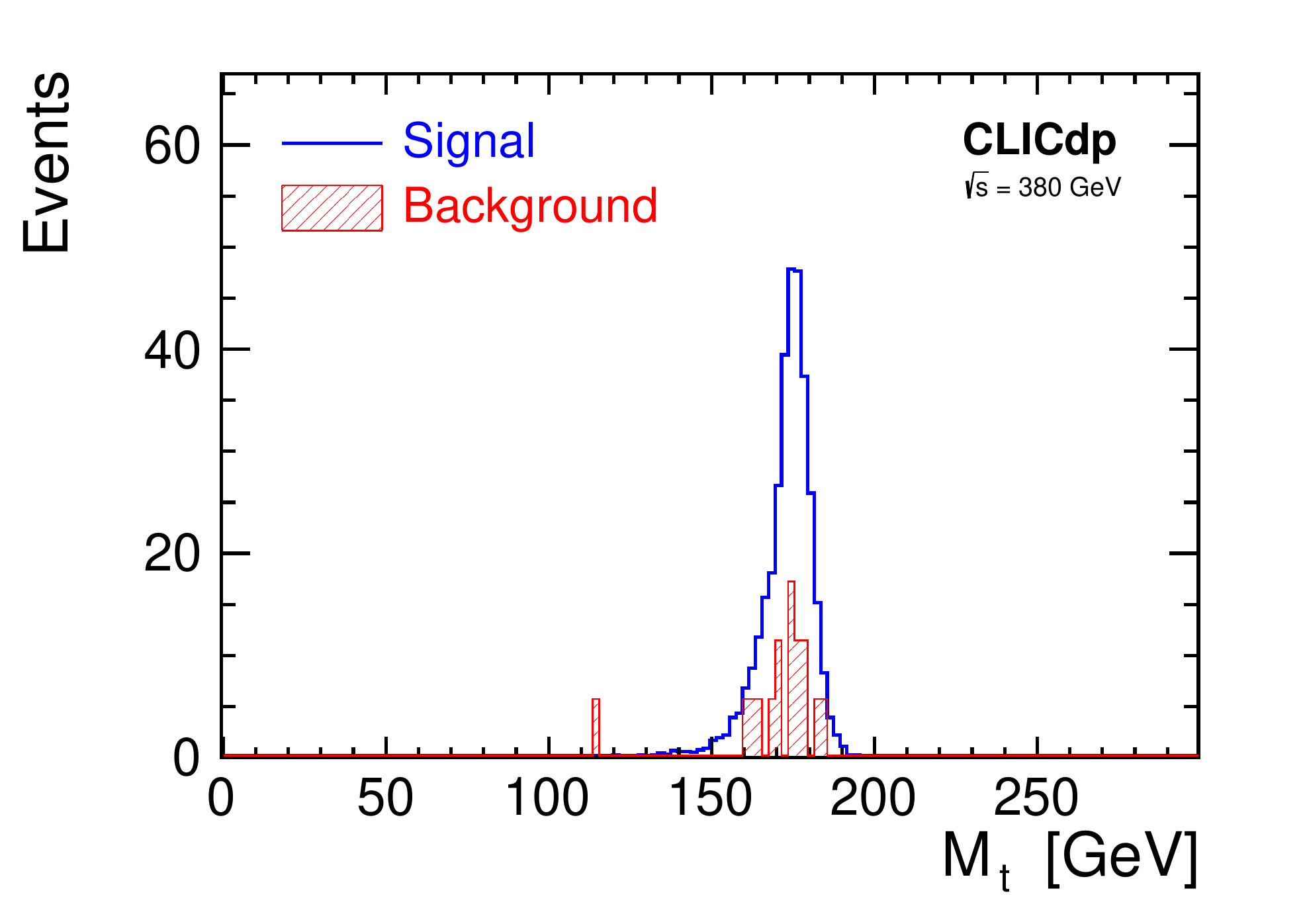}
\includegraphics[height=5cm]{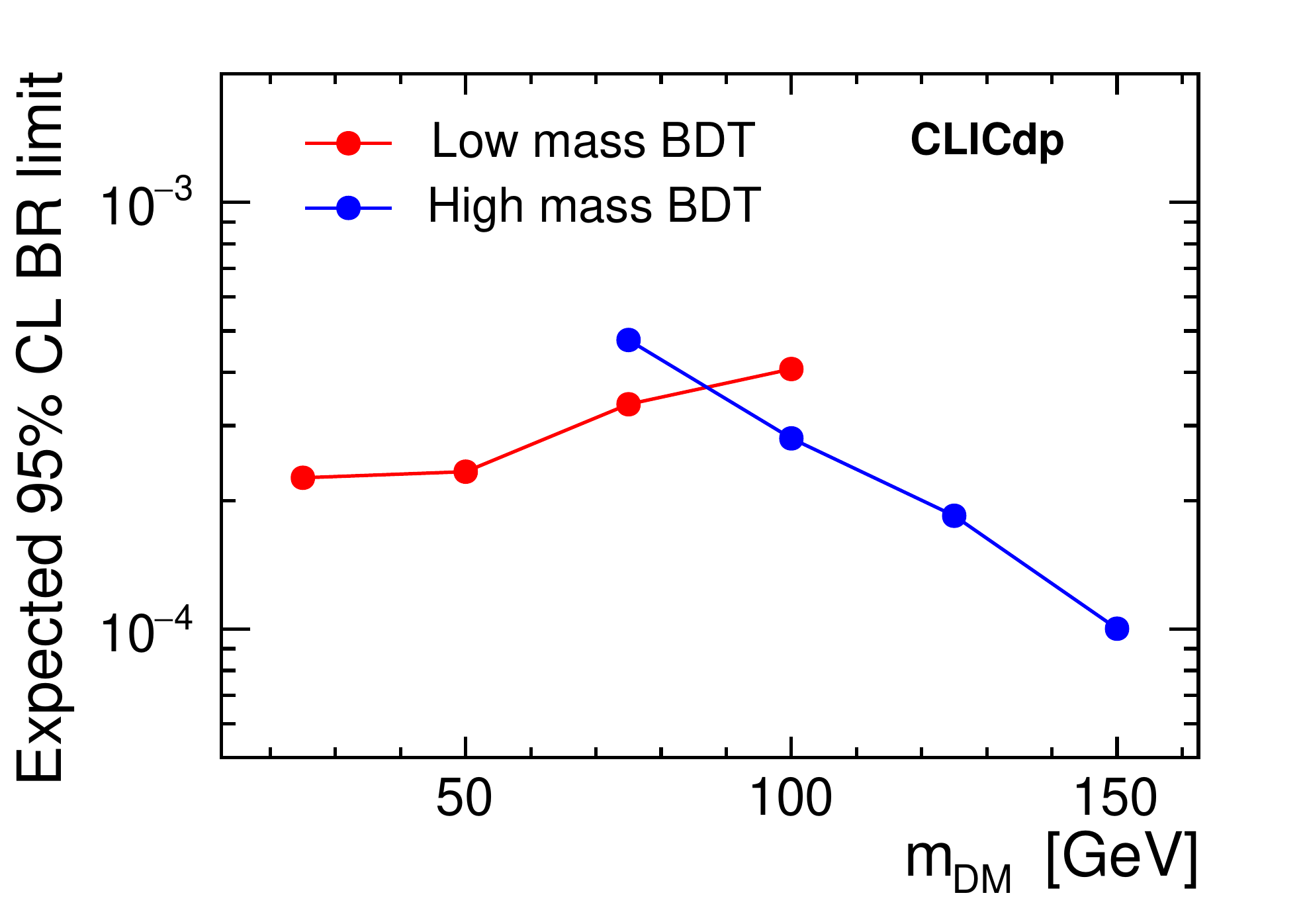}
\end{center}
\vspace*{-0.5cm}
  \caption{Left: invariant mass distribution of the top quark from the
    FCNC decay \tcg reconstructed at 380 GeV CLIC after selection
    based on the BDT response, normalised to 1.0\,ab$^{-1}$ and
    BR(\tcg) $=10^{-3}$ for the signal events.
    Right: limits at 95\% C.L. on the top quark FCNC decay \tcx
    expected for 1.0\,ab$^{-1}$ collected at 380\,GeV CLIC \cite{1807.02441}.}  
  \label{fig:fcnc}
\end{figure}

Flavour-Changing Neutral Current (FCNC) top quark decays are
very strongly suppressed in the Standard Model, with the expected
branching ratios
BR$(\ttbar \to \PQc$X$)   \sim   10^{-15}$ to $10^{-12}$
(X$ = \PGg, \; \Pg, \; \PZ, \; \PH$).
Observation of any such decay would be a direct signature
for ``new physics''.
Significant enhancement is expected in many BSM
scenarios, reaching up to $10^{-2}$ for BR(\tch) and $10^{-5}$ for BR(\tcg).
Three FCNC decay channels involving charm quark have been studied for
the first stage of CLIC. 
All channels profit from the precise final state reconstruction and
high flavour tagging efficiency.
Selected results are shown in Fig.~\ref{fig:fcnc}. 
Limits at 95\% C.L. expected for 1\,ab$^{-1}$ collected at 380\,GeV
CLIC are \cite{1807.02441}: 
\begin{eqnarray*}
  \text{BR}(\tcg)  & < & 2.6 \cdot 10^{-5}\, ,  \\
  \text{BR}(\tch) \times 
  \text{BR}(\hbb)  & < & 8.8 \cdot 10^{-5}\, , \\
 \text{BR}(\tcx)  & < & 1.0 - 3.4 \cdot 10^{-4} \, , 
 \end{eqnarray*}
where the limit for the top-quark decay involving invisible massive
scalar particle, \tcx, depends on the assumed particle mass, see
Fig.~\ref{fig:fcnc} (right). 
For channels involving charm quark, limits expected at ILC and CLIC are
much stronger than those estimated for HL-LHC \cite{fcnc_cms,fcnc_atlas}.


\subsection{Constraining top-quark couplings}

\begin{figure}[tb]
\begin{center}
   \includegraphics[height=6cm]{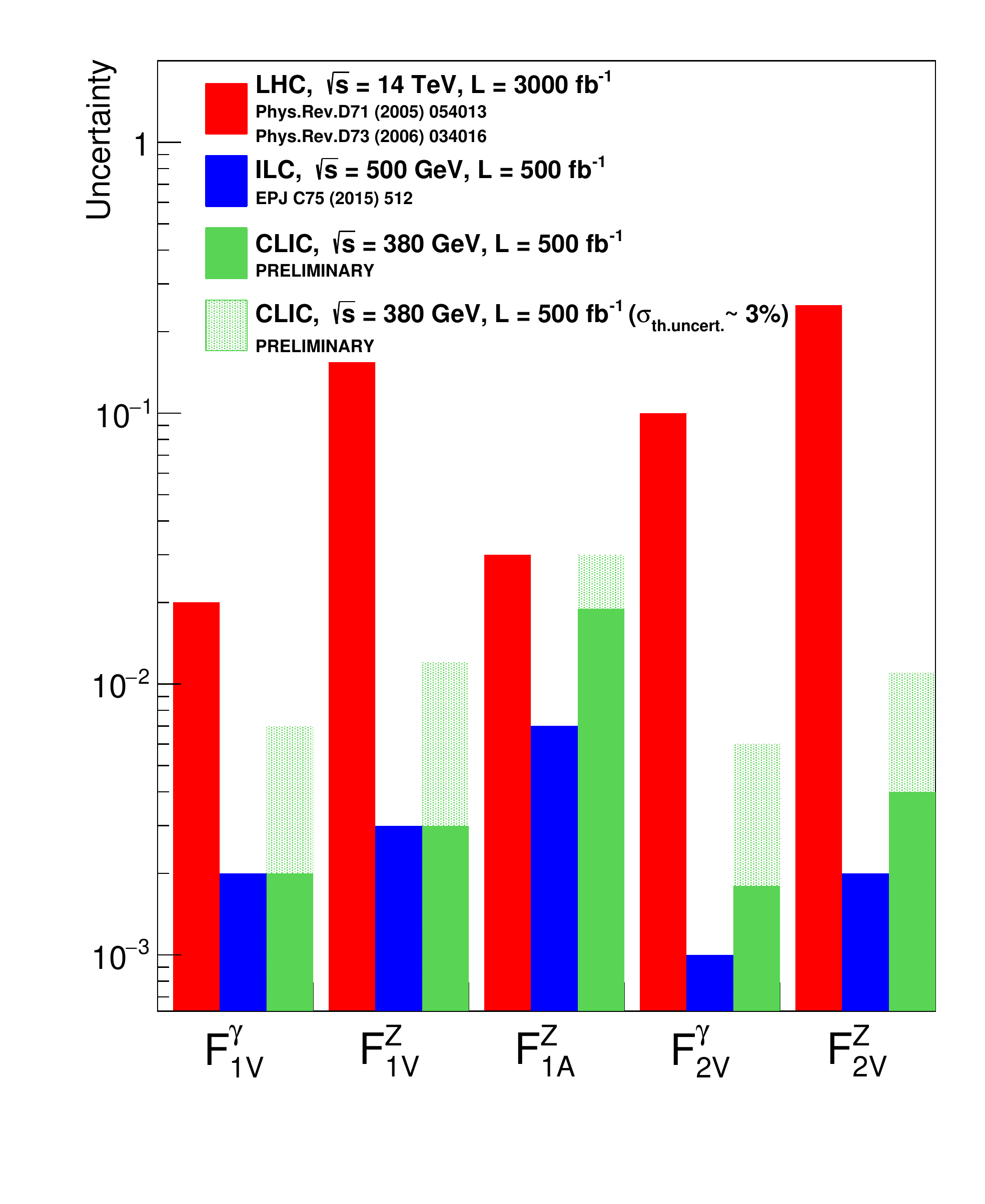}
   \includegraphics[height=6cm]{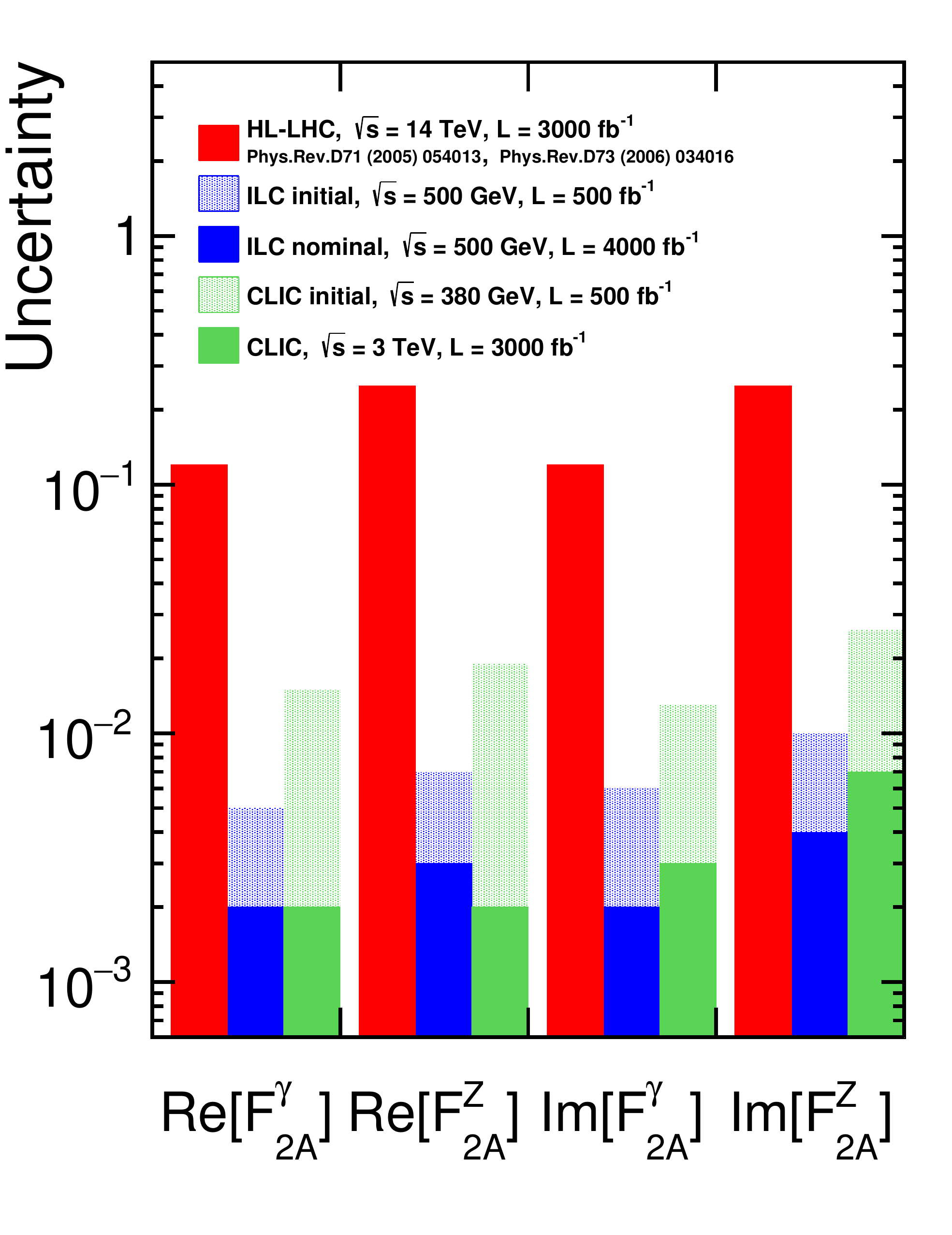}
\end{center}
\vspace*{-0.5cm}
  \caption{Comparison of the uncertainties on CP-conserving (left) and
    CP-violating (right) form factors in top-quark electroweak
    couplings expected at the HL-LHC, ILC and CLIC \cite{1710.06737}.}   
  \label{fig:tew}
\end{figure}

Measurements of the \ttbar production cross section and angular
distributions for different centre-of-mass energies and beam polarisations
can be used for precise determination of the top-quark couplings to
the photon and the Z-boson, and to constrain possible BSM effects. 
The expected sensitivity of ILC and CLIC to electroweak couplings of the
top quark is presented in Fig.~\ref{fig:tew} in terms of the
uncertainties on the CP-conserving and CP-violating form factors
\cite{1710.06737}. 

\begin{figure}[tb]
\begin{center}
   \includegraphics[height=6cm]{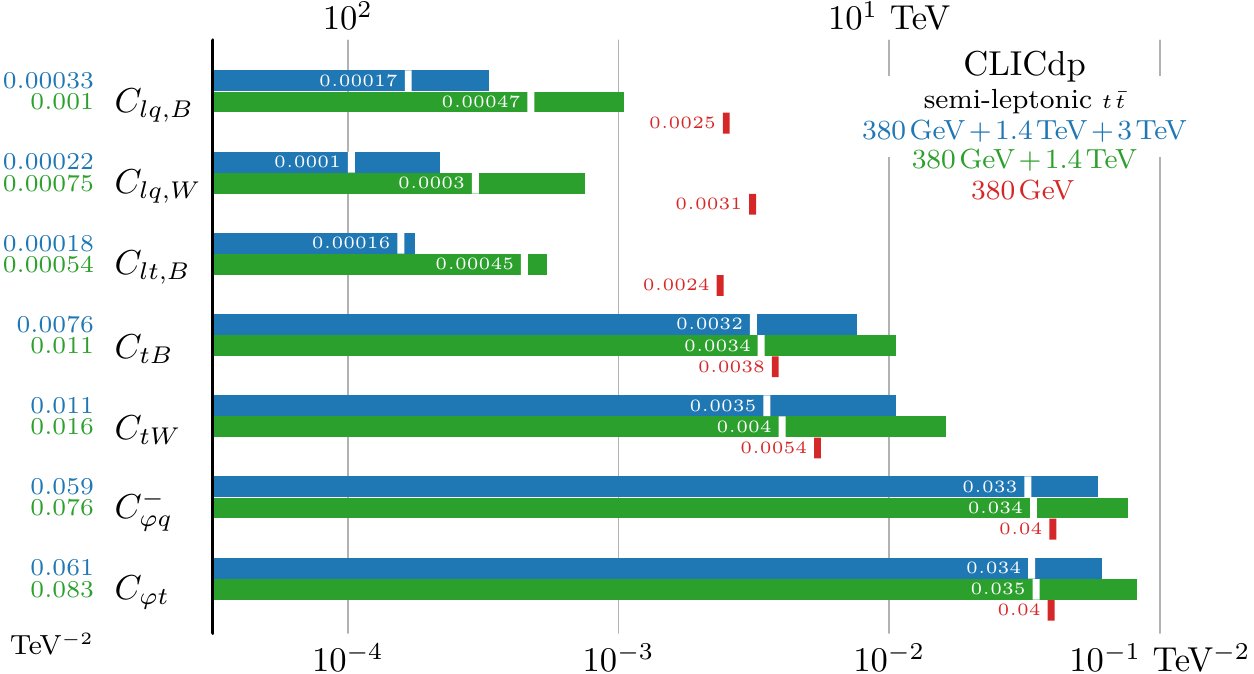}
\end{center}
\vspace*{-0.5cm}
  \caption{Summary of the global EFT analysis results using
    statistically optimal observables for the three CLIC energy
    stages. The colour bars indicate the 1$\sigma$ constraints on each
    of the seven Wilson coefficients. The corresponding individual
    operator sensitivities are shown as ticks \cite{1807.02441}.} 
  \label{fig:teft}
\end{figure}

Effects induced by heavy new physics can be also described in terms of
Effective Field Theory (EFT) operators.
Expected limits on the Wilson coefficients for seven EFT operators
contributing to the top-quark pair production, resulting from the
global EFT analysis of all measurements based on  statistically
optimal observables,  are presented in Fig.~\ref{fig:teft}.
Even at the first CLIC stage, mass scales in the 10\,TeV range can
be probed. For four operators, most of the sensitivity is provided by 
the initial 380\,GeV stage \cite{1807.02441}.
These results confirm that high energy CLIC can reach ``new physics''
scales in the 100\,TeV domain.


\section{BSM physics}

Strong limits on many BSM scenarios are expected from the experiments
at HL-LHC. 
However, searches which can be performed at \epem colliders are 
in many cases complementary.
Two approaches are possible:  direct BSM searches, in particular for
models with weak couplings or soft signatures, which are difficult to
be constrained at the LHC and indirect searches, where the sensitivity
to ``new physics'' effects can reach very high energy scales thanks to high
measurement precision and clean environment. 

 
\subsection{Search for new scalars}

\begin{figure}[tb]
\begin{center}
\includegraphics[height=5cm]{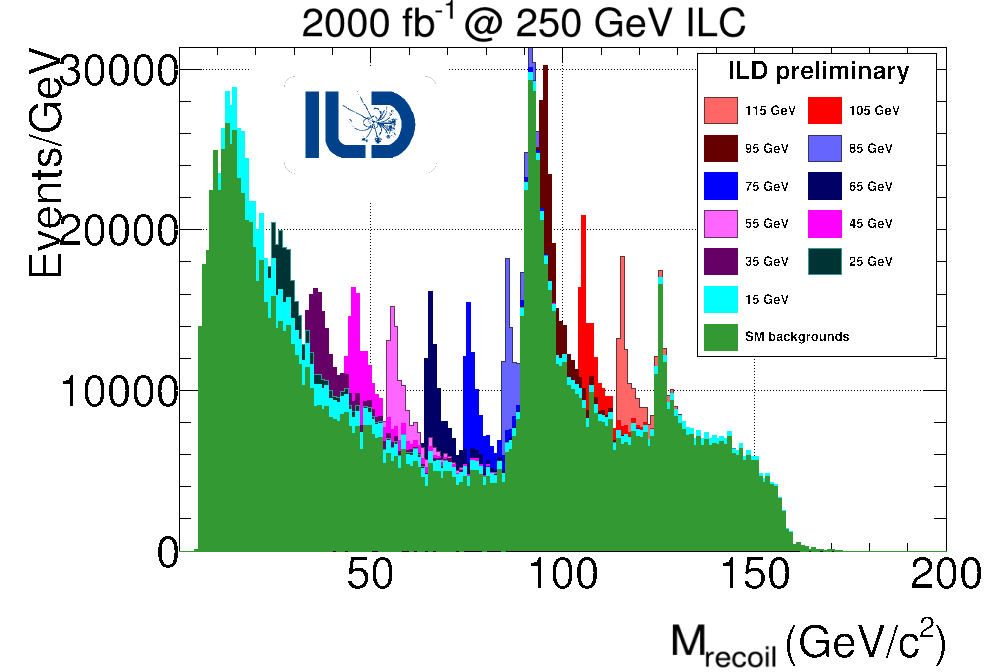}
\end{center}
\vspace*{-0.5cm}
  \caption{The recoil mass distributions for signals of new scalar
    production with different masses and for all SM backgrounds, after
    the selection cuts, for ILC running at 250\,GeV \cite{ilchinv}. } 
  \label{fig:hinvrec}
\end{figure}

\begin{figure}[t]
\begin{center}
\includegraphics[height=5cm]{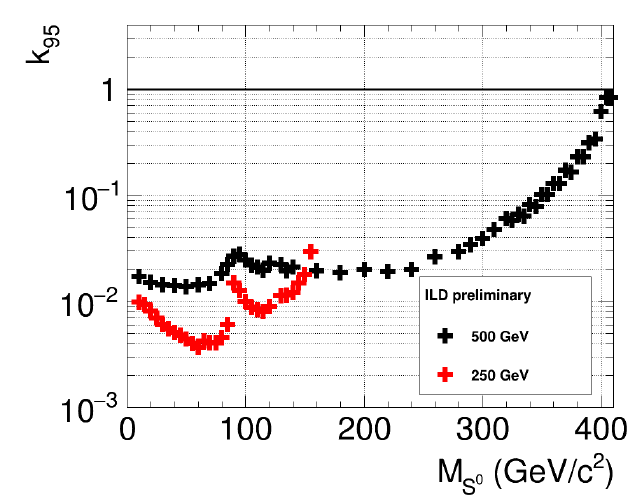}
\includegraphics[height=4.7cm]{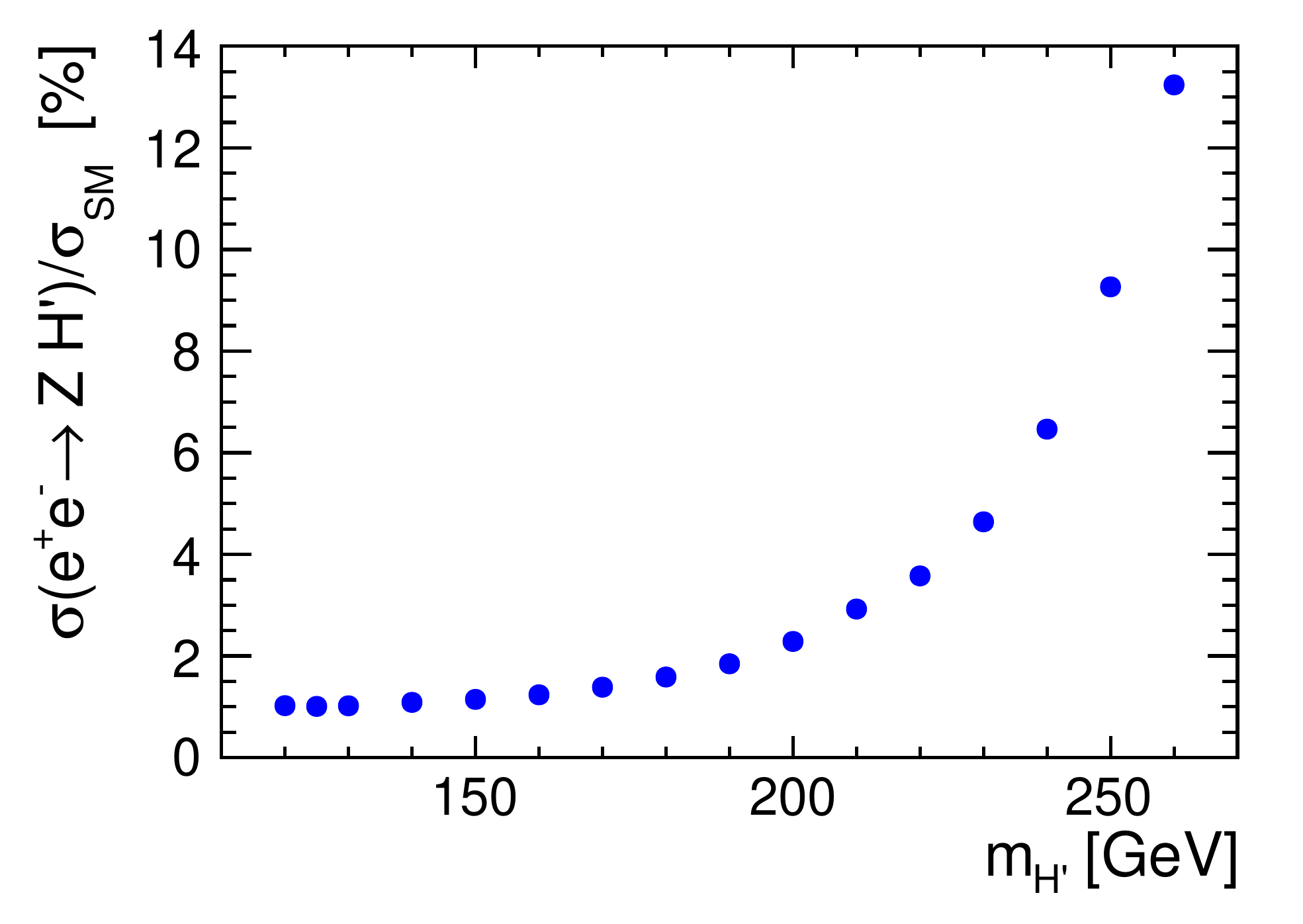}\mylabel{4.5cm}{4.7cm}{CLICdp}
\end{center}
\vspace*{-0.5cm}
  \caption{Expected limits on the production cross section of the new
    scalar, relative to the expected SM Higgs production cross
    section, as a function of its mass, for ILC running at 250\,GeV
    and 500\,GeV \cite{ilchinv} (left) and CLIC running at 380\,GeV 
\cite{2002.06034} (right).
} 
  \label{fig:hinvlim}
\end{figure}

Many BSM models introduce extended Higgs sectors.
If couplings of new scalars to SM particles are small,
they are not excluded by the current experimental constraints, even
for masses of the order of 10--100\,GeV.
Such a new scalar, $\HepParticle{S}{}{0}$, if coupling to the SM gauge
bosons, could be produced in the ``Scalarstrahlung'' process,
$\Pep\Pem \to \PZ \HepParticle{S}{}{0}$. 
With very weak couplings to SM particles, invisible decays of the new
scalar are expected to dominate.
Such decays can be constrained by looking at events with single
Z-boson production and searching for a peak in the recoil mass
distribution (similar to the search for invisible Higgs boson decays
presented above).
An example of the reconstructed recoil mass distribution for SM
background processes and new scalar production with different
masses, for ILC running at 250\,GeV, is presented in Fig.~\ref{fig:hinvrec}.
As the recoil mass can be very precisely reconstructed for Z-boson
decays into two muons,  new scalar production should be clearly
visible as an additional structure in the recoil mass distribution. 
Resulting limits on the production cross section of the new scalar,
relative to the expected SM Higgs production cross section at given
mass, estimated for ILC running at 250\,GeV and 500\,GeV
\cite{ilchinv} are shown in Fig.~\ref{fig:hinvlim} (left). 
In the study performed for CLIC running at 380\,GeV \cite{2002.06034},
see Fig.~\ref{fig:hinvlim} (right), hadronic decays of the Z-boson
were considered and the new scalar was assumed to have only invisible
decay channels.   
While ILC running at 250\,GeV is expected to be most sensitive to
production of light new scalars below 150\,GeV, running at higher
centre-of-mass energy  allows to extend exclusion limits to higher
scalar masses.

\subsection{Inert Doublet Model}

The Inert Doublet Model is one of the simplest SM extensions
providing a natural candidate for dark matter.
Light IDM scenarios, with scalar masses in $\mathcal{O}$(100\,GeV)
range are still not excluded by the current experimental and
theoretical constraints \cite{Kalinowski:2018ylg}. 
Low mass IDM scenarios can be observed with high significance in the
di-lepton channels already at a \epem collider with 250\,GeV
centre-of-mass energy~\cite{Zarnecki:2019poj}.
The discovery reach increases for higher $\sqrt{s}$
\cite{Kalinowski:2018kdn} and significant improvement in the discovery
reach is observed when considering the semi-leptonic final state
\cite{Sokolowska:2019xhe,2002.11716}.  
In particular, the discovery reach for charged scalar
pair-production is extended at 3\,TeV CLIC up to $m_{\PH^\pm} \sim$
1\,TeV. 
A full simulation study of the charge scalar pair-production in the
semi-leptonic decay channel is ongoing.
More details were given in a dedicated contribution \cite{2002.11716}.


 \subsection{Dark Matter searches}

\begin{figure}[tb]
\begin{center}
 \includegraphics[height=5cm]{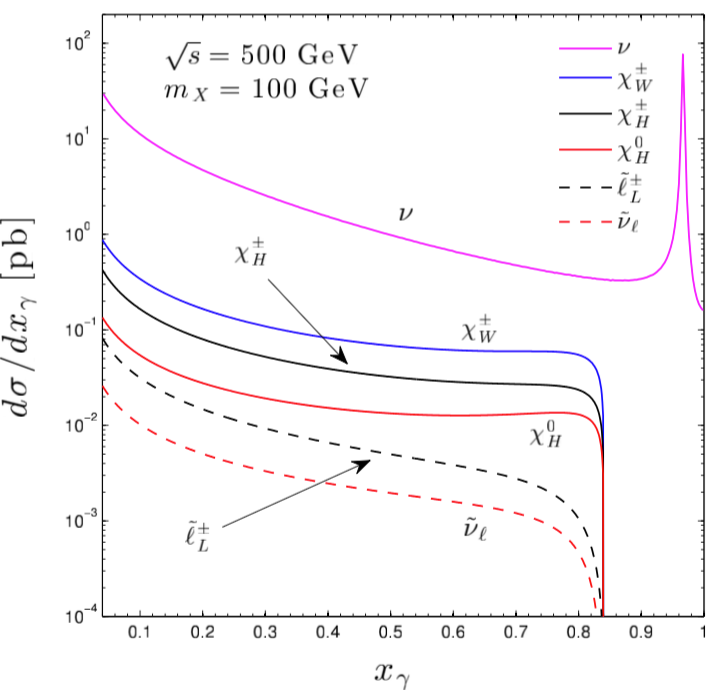}
 \includegraphics[height=5cm]{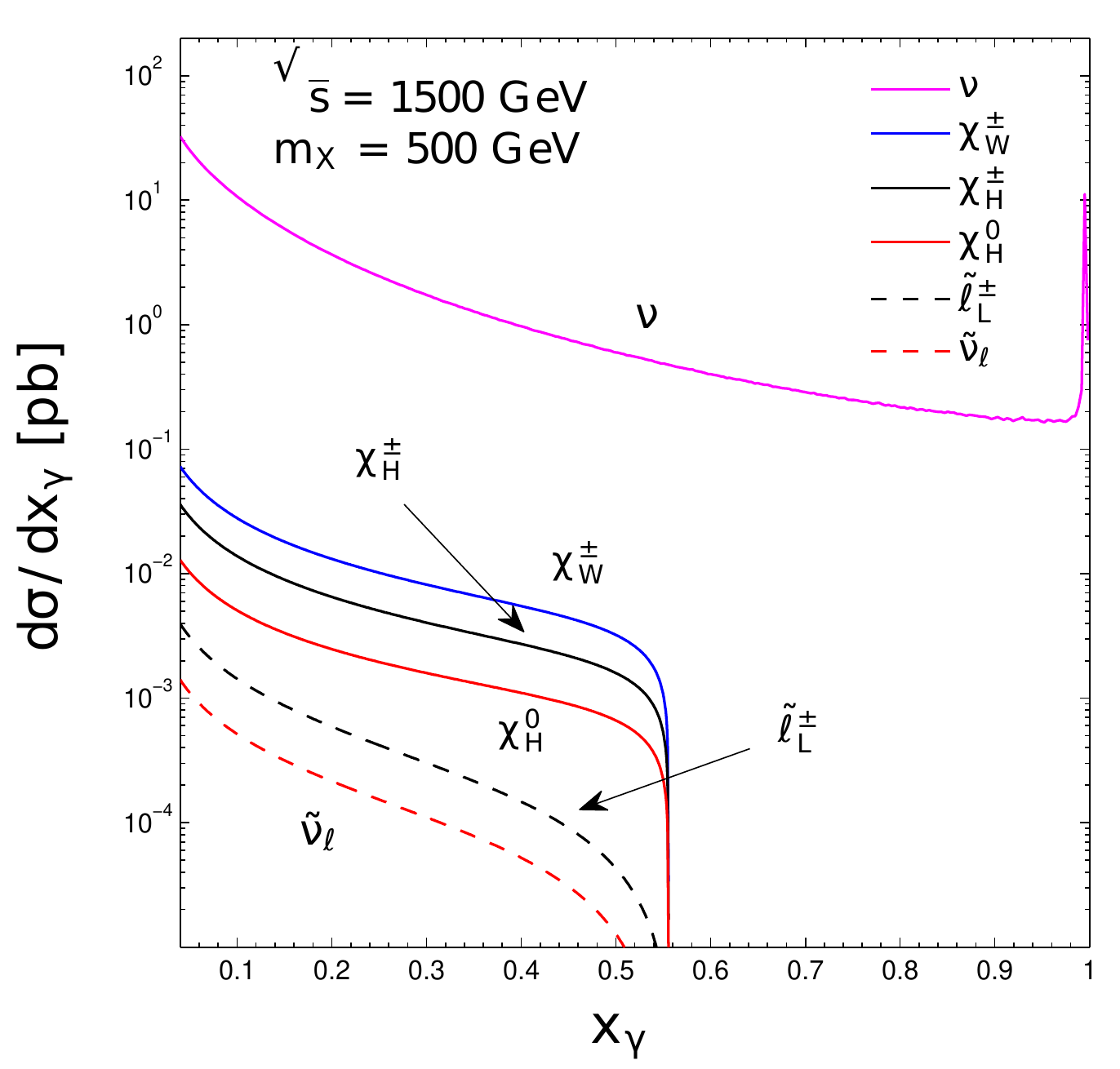}
\end{center}
\vspace*{-0.5cm}
  \caption{Unpolarized $x_\gamma$ distribution
    $\frac{d\sigma}{dx_\gamma}$ for different SUSY EW particles, as
    well as that of the background process $\Pep\Pem\to\PGn\PAGn$
    (solid line on the top), for $m_X = 100$\,GeV at a 500 GeV ILC \cite{Choi:2015zka}
    (left) and  $m_X = 500$\,GeV at a 1500 GeV CLIC \cite{JKclic2016} (right).
  } 
  \label{fig:mphth}
\end{figure}

\begin{figure}[t]
\begin{center}
  \includegraphics[height=5cm]{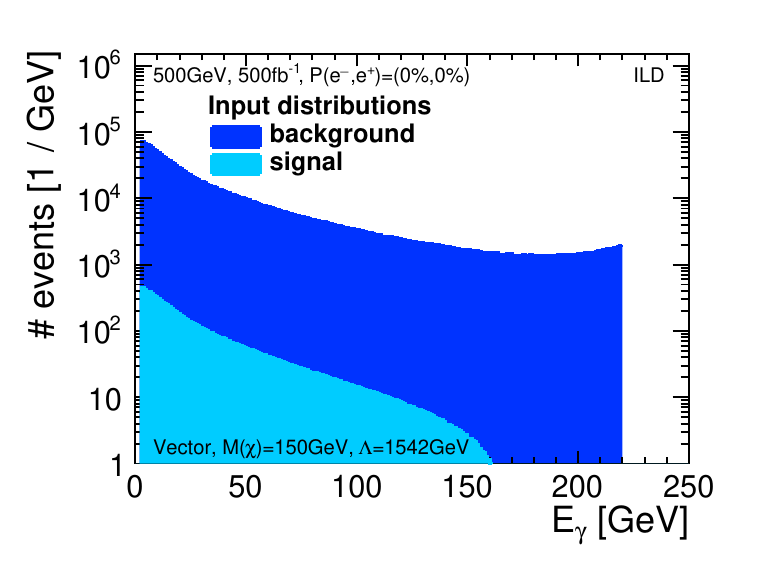}
  \includegraphics[height=5cm]{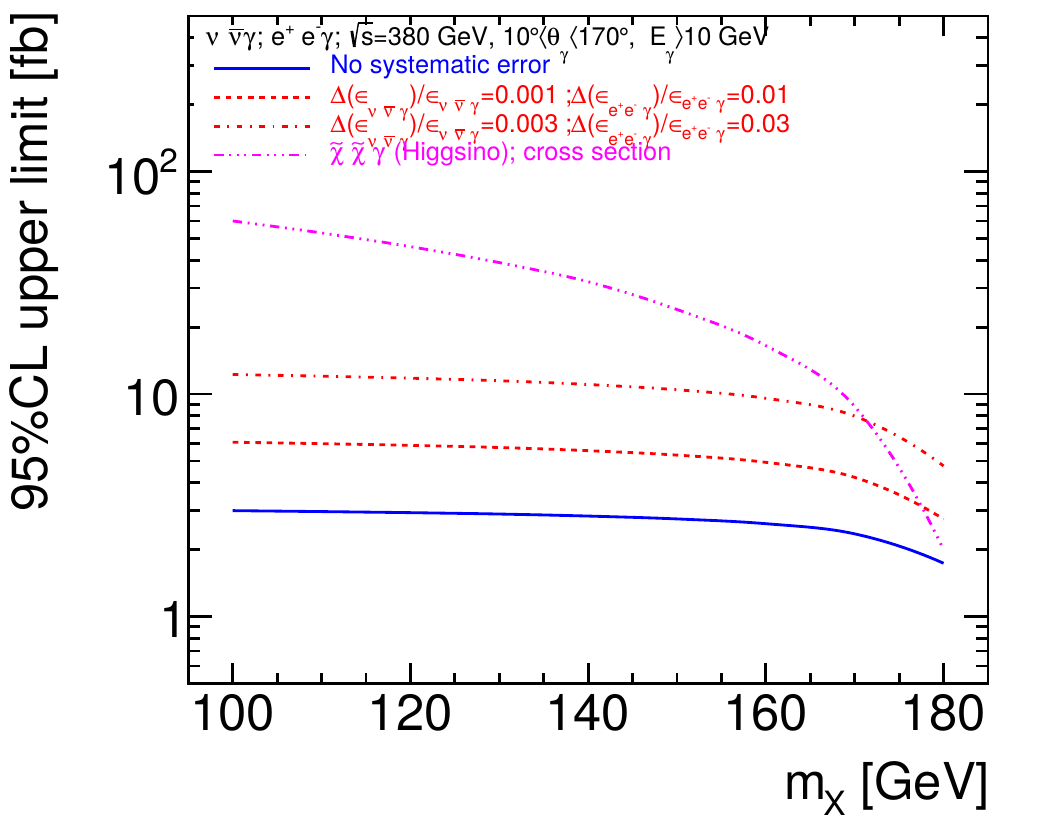}
\end{center}
\vspace*{-0.5cm}
  \caption{Left: the signal and background input histograms of the photon energy
    distribution, for pair-production of DM particles with mass of
    150\,GeV, corresponding to the expected 95\% C.L. exclusion limit
    for the production  cross section at 500\,GeV ILC
    \cite{Habermehl:2018yul}.  
   Right: upper limit (95\% C.L.) on the $\Pep\Pem \to \PGc\PGc\PGg$
   cross section at $\sqrt{s} = 380$\,GeV as a function of 
  the DM particle mass, $m_\chi$. The limits without and with
  including systematic uncertainties are compared to the expected
  cross section for the higgsino pair production with an ISR photon
  \cite{1812.02093}.  
  } 
  \label{fig:mphex}
\end{figure}

\begin{figure}[p]
\begin{center}
  \includegraphics[height=6cm]{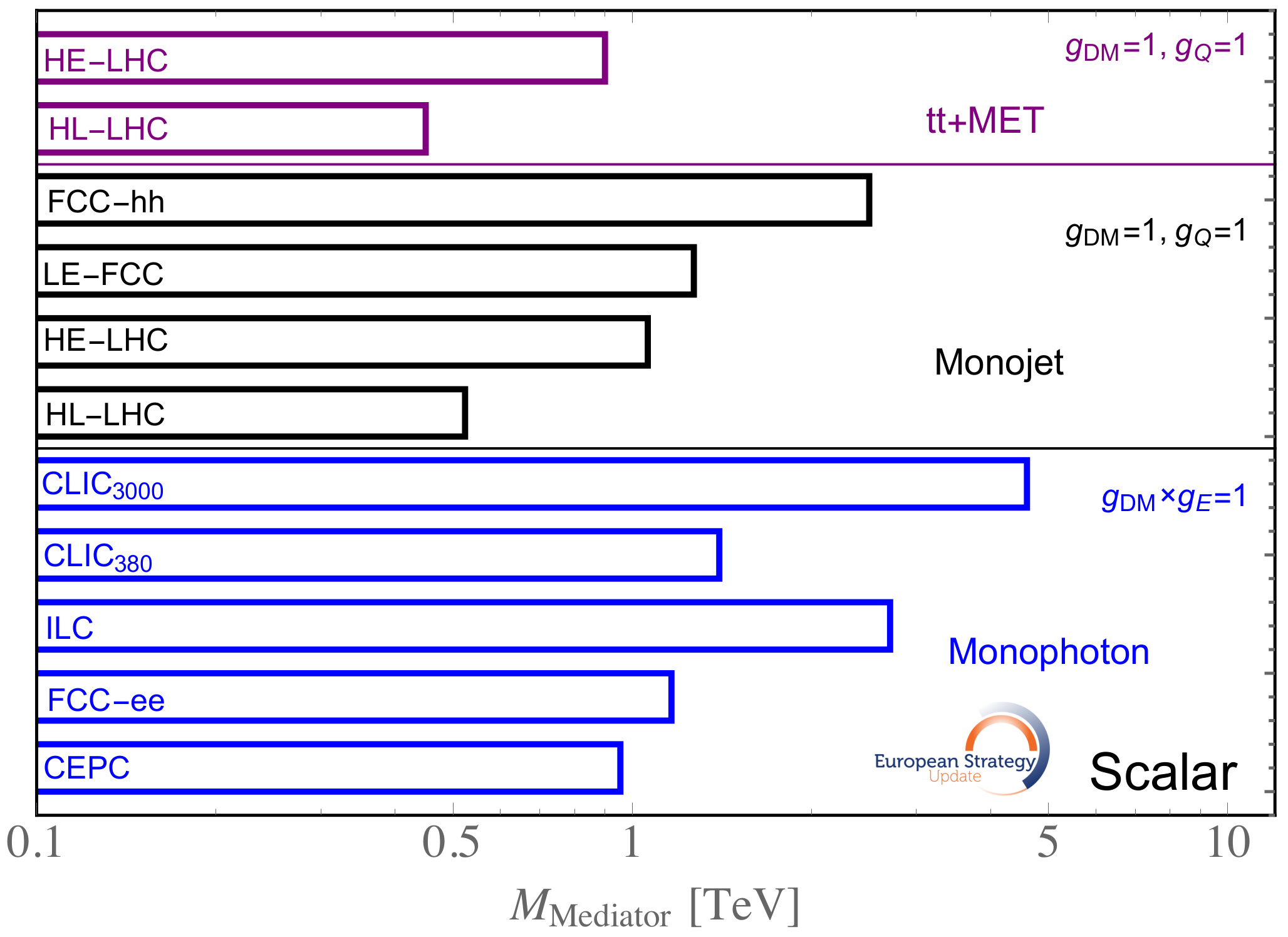}
\end{center}
\vspace*{-0.5cm}
  \caption{Summary of 2$\sigma$ sensitivity to scalar mediator mass
    at future colliders for simplified models with a DM mass of
    $M_{DM} = 1$\,GeV and for the couplings shown in the figure
    \cite{Strategy:2019vxc}.}   
  \label{fig:mphcomp}
\end{figure}

Production of Dark Matter particles in \epem collisions is possible in
many BSM scenarios. 
The unique advantage of \epem colliders is that we can detect
processes with  invisible final states by studying the spectra of ISR
photons.
Expected spectra of the photon energy (relative to the beam energy:
$x_\gamma = E_\gamma / E_e$) is presented in Fig.~\ref{fig:mphth} for
production of different SUSY particles and the SM background process
$\Pep\Pem\to\PGn\PAGn\PGg$, for ILC running at 500\,GeV
\cite{Choi:2015zka} and 1.5\,TeV CLIC \cite{JKclic2016}. 
A full simulation study performed for the ILD
\cite{Habermehl:2018yul,2001.03011} indicates that detector effects
are very important and the background levels are high, see
Fig.~\ref{fig:mphex} (left).
The $\PGn\PAGn\PGg$ background is strongly suppressed for the 
right-handed electron beam polarisation and reduction by up to a factor of
5 is expected at the ILC thanks to the positron beam polarisation
\cite{2001.03011}. 
Comparison of the measured photon energy spectra with SM expectations
can be used to set limits on the cross section contribution from the
pair-production of DM particles, as shown in Fig.~\ref{fig:mphex}
(right) for CLIC running at 380\,GeV \cite{1812.02093}.
Comparison of extracted mediator mass limits, for mono-photon events
at future \epem colliders and mono-jet events at future hadronic
colliders, is presented in Fig.~\ref{fig:mphcomp}.
For scalar mediator hypothesis, mediator mass limits expected at 
ILC and CLIC are comparable with those at FCC-hh \cite{Strategy:2019vxc}.


\subsection{Other direct searches}

\begin{figure}[p]
\begin{center}
  \includegraphics[height=5cm]{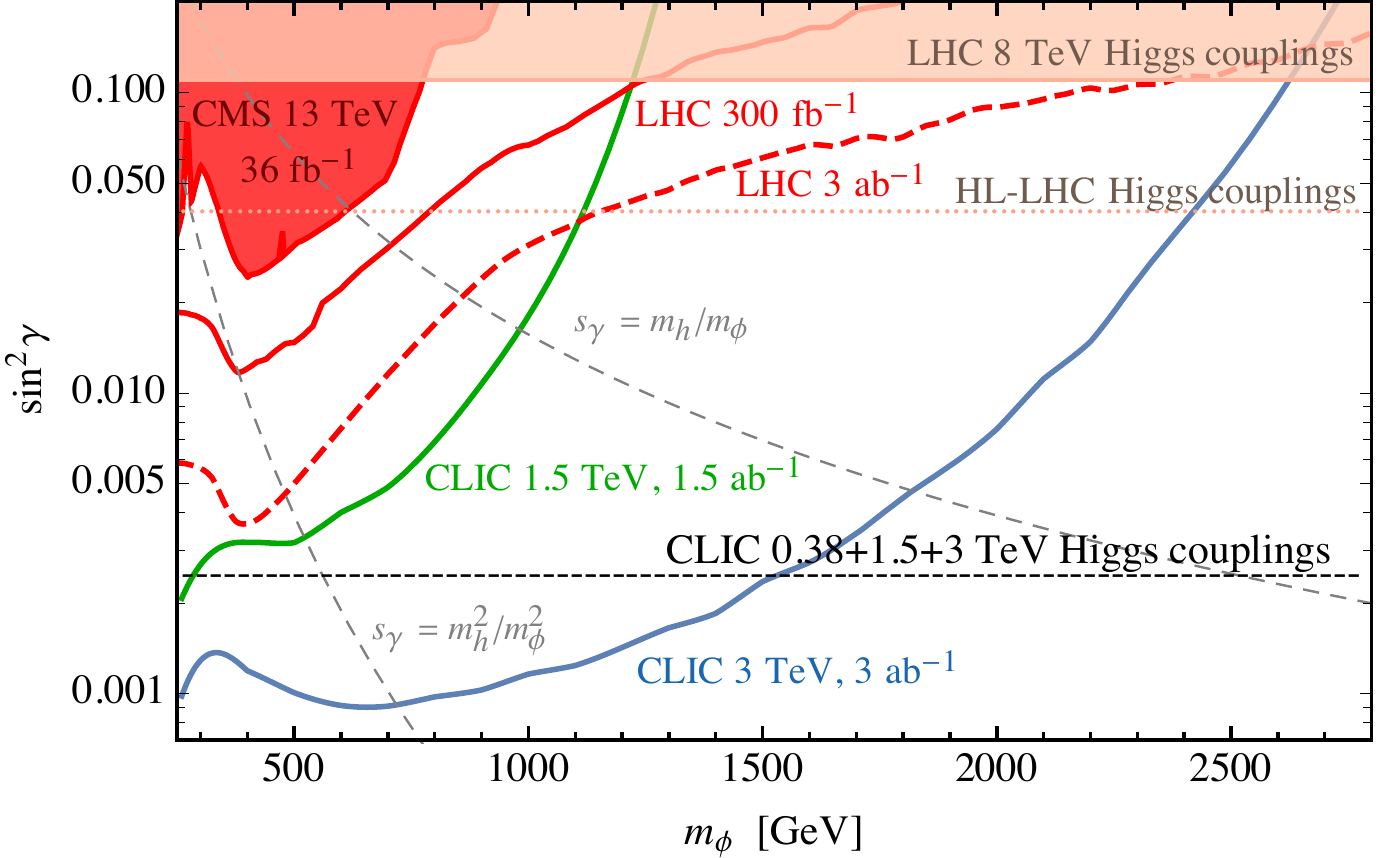}
\end{center}
\vspace*{-0.5cm}
  \caption{New scalar singlet constraints at 95\% C.L. in the plane
    $(m_\phi , \sin 2\gamma)$.
    Direct and indirect limits estimated for CLIC are compared to
    those expected at HL-LHC.
    The shaded regions are the present constraints from LHC. 
    From \cite{1812.02093}.}   
  \label{fig:singlet}
\end{figure}

\begin{figure}[p]
\begin{center}
   \includegraphics[height=5cm]{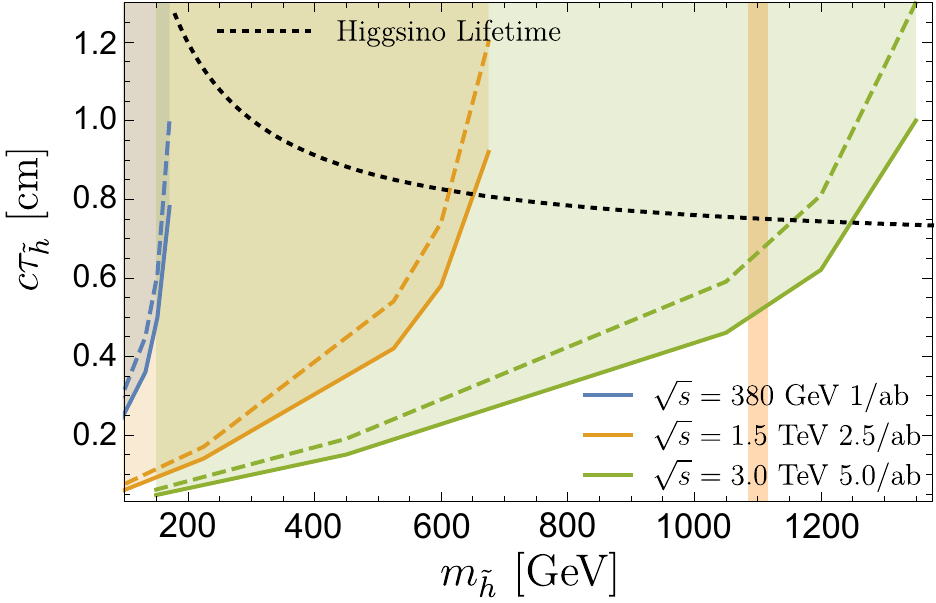}
\end{center}
\vspace*{-0.5cm}
  \caption{Contours in the lifetime-mass plane for N=3 (solid) and
    N=30 (dashed) higgsino events in the detector acceptance range at
    the three stages of CLIC \cite{1812.06018}.
    The black dashed line indicates the lifetime of the pure Higgsino
    state of a given mass expected for the scenario considered in
    \cite{1812.02093}.}     
  \label{fig:higgsino}
\end{figure}

The direct (and indirect) reach of ILC and CLIC can exceed that of HL-LHC
for many BSM models, in particular those with exotic scalar sector
or new Higgs bosons.
Shown in Fig.~\ref{fig:singlet} are direct and indirect sensitivities
of CLIC to new heavy scalar singlets, compared to those expected at
HL-LHC \cite{1812.02093}.  
Constraints at 95\% C.L. are presented in the plane of scalar mass,
$m_\phi$, and the mixing angle of new scalar field with the SM field,
$\sin 2\gamma$.
The results presented demonstrate that direct and indirect limits from
CLIC are complementary and much stronger than the expected HL-LHC
sensitivity. 
High sensitivity, thanks to precision tracking and low background
conditions, is also expected in searches based on the ``disappearing
tracks'' signature.
Results indicating CLIC sensitivity to charged higgsino pair
production are presented in Fig.~\ref{fig:higgsino} \cite{1812.06018}.

  \subsection{EFT analysis}

\begin{figure}[tb]
\begin{center}
\includegraphics[height=6cm]{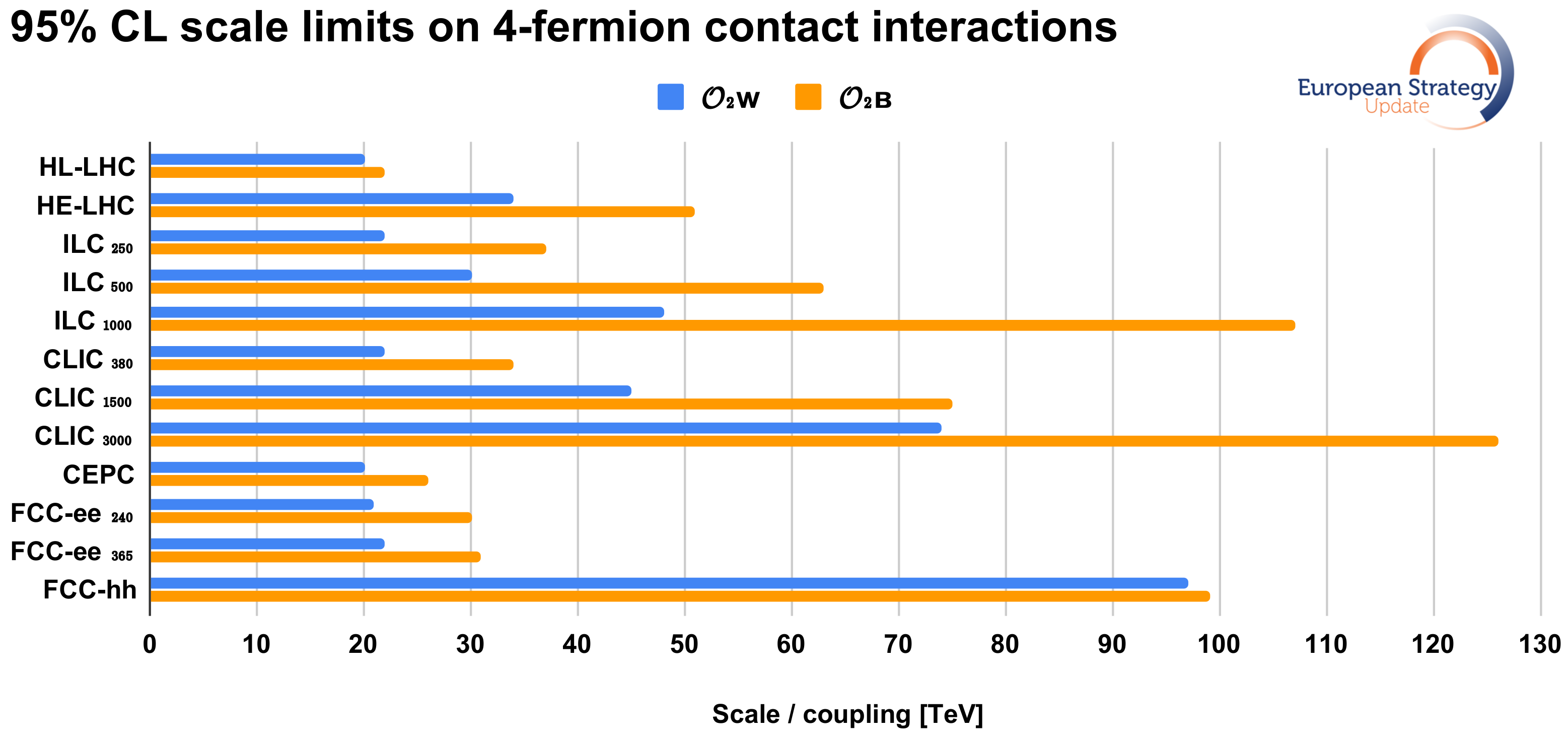}
\includegraphics[height=6cm]{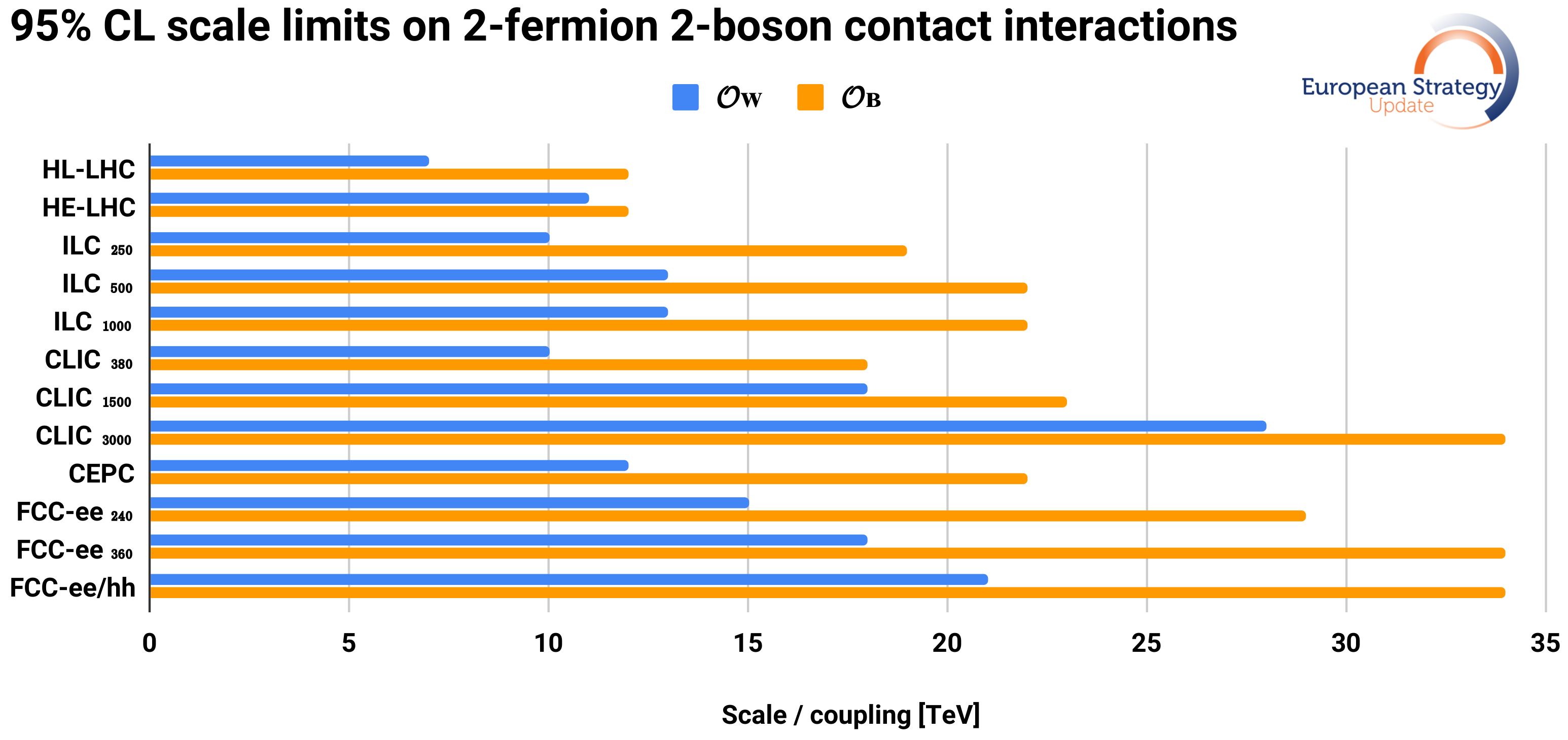}
\end{center}
\vspace*{-0.5cm}
  \caption{Exclusion reach of different colliders on: (upper plot)
    four-fermion contact interactions from the operators
    $\mathcal{O}_{2W}$ and $\mathcal{O}_{2B}$,
    and (bottom plot) two-fermion/two-boson contact interactions from
    the operators $\mathcal{O}_{W}$ and $\mathcal{O}_{B}$ \cite{Strategy:2019vxc}.}   
  \label{fig:ci}
\end{figure}

All collider measurements can be combined in a global EFT analysis
which gives the most general constraints on the possible ``new
physics'' effects at high energy scales.
A summary of the sensitivity expected at different future colliders is
presented in Fig.~\ref{fig:ci}, for 4-fermion contact interactions
and 2-fermion-2-boson contact interaction operators \cite{Strategy:2019vxc}.
The ILC and CLIC sensitivities exceed that of HL-LHC even at their first
running stages.
For 4-fermion contact interactions limits reaching 100\,TeV scales,
can only be obtained at (upgraded) 1\,TeV ILC, CLIC running at 3\,TeV
or FCC-hh.


\section{Conclusions}

Presented in this contribution is my personal selection of results
demonstrating the physics potential of ILC and CLIC.
High Energy linear \epem colliders offer a rich and diverse research
programme: 
 precise determination of Higgs couplings;
 precise determination of top-quark mass and other properties;
 stringent constraints on many BSM scenarios from indirect searches
 reaching mass scales up to 100\,TeV
 and prospects for direct observation of new physics in many scenarios.
Because of the different energy choices and running scenarios, the two
projects, ILC and CLIC, are to a large extent complementary. 
From the pure physics point of view, in would be most
advantageous, if both colliders are built.

\subsection*{Acknowledgements}

I would like to thank ILD, SiD and CLICdp groups for providing results
presented in this contribution, as well as the group members for many
useful comments and suggestions.
This contribution was supported by the National Science Centre, Poland, the
OPUS project under contract UMO-2017/25/B/ST2/00496 (2018-2021).
Thanks are due to the LCC generator working group
and the ILD software working group for providing the simulation and
reconstruction tools and producing the Monte Carlo samples.
Presented studies benefited also from computing services provided
by the ILC Virtual  Organization, supported by the national resource
providers of the EGI Federation and the Open Science GRID.

\end{document}